\documentclass[acmsmall,nonacm]{acmart}

\AtBeginDocument{%
  }

\renewcommand\footnotetextcopyrightpermission[1]{}
\makeatletter
\let\@authorsaddresses\@empty
\makeatother

\usepackage{tikz}
\usepackage{braket}
\usepackage{lipsum}
\usepackage{subfig}
\usepackage{xcolor}
\usepackage{amsmath}
\usepackage{balance}
\usepackage{environ}
\usepackage{physics}
\usepackage{colortbl}
\usepackage{enumitem}
\usepackage{amsfonts}
\usepackage{booktabs}
\usepackage{graphicx}
\usepackage{textcomp}
\usepackage{wrapfig}
\usepackage{algorithm}
\usepackage{algpseudocode}
\usepackage{arydshln}
\usepackage[normalem]{ulem}
\usepackage{multirow}
\usepackage{booktabs}
\usepackage{multirow}
\usepackage{colortbl}
\usepackage{tikz}
\usepackage{makecell}
\usepackage{multicol}
\usepackage[most]{tcolorbox}

\definecolor{qoncord}{HTML}{ffad31}
\definecolor{bestmap}{HTML}{1e932e}
\definecolor{nest}{HTML}{00185f}

\newcommand{\sol}{NEST}
\newcommand{\heh}{HeH$^+$}
\newcommand{\htwo}{H$_2$}
\newcommand{\hthree}{H$_3^+$}

\newcommand{\rev}[1]{\textcolor{black}{#1}}

\begin{document}

\title{Three Birds with One Stone: Improving Performance, Convergence, and System Throughput with \sol{}}

\author{Yuqian Huo}\affiliation{\institution{Rice University}\country{Houston, TX, USA}}
\author{David Quiroga}\affiliation{\institution{Rice University}\country{Houston, TX, USA}}
\author{Anastasios Kyrillidis}\affiliation{\institution{Rice University}\country{Houston, TX, USA}}
\author{Tirthak Patel}\affiliation{\institution{Rice University}\country{Houston, TX, USA}}

\renewcommand{\shortauthors}{Yuqian Huo, David Quiroga, Anastasios Kyrillidis, Tirthak Patel}
    
\begin{abstract}

Variational quantum algorithms (VQAs) have the potential to demonstrate quantum utility on near-term quantum computers. However, these algorithms often get executed on the highest-fidelity qubits and computers to achieve the best performance, causing low system throughput. Recent efforts have shown that VQAs can be run on low-fidelity qubits initially and high-fidelity qubits later on to still achieve good performance. We take this effort forward and show that carefully varying the qubit fidelity map of the VQA over its execution using our technique, \sol{}, does not just (1) improve performance (i.e., help achieve close to optimal results), but also (2) lead to faster convergence. We also use \sol{} to co-locate multiple VQAs concurrently on the same computer, thus (3) increasing the system throughput, and therefore, balancing and optimizing three conflicting metrics simultaneously.

\end{abstract}

\begin{CCSXML}
<ccs2012>
   <concept>
       <concept_id>10010520.10010521.10010542.10010550</concept_id>
       <concept_desc>Computer systems organization~Quantum computing</concept_desc>
       <concept_significance>500</concept_significance>
   </concept>
 </ccs2012>
\end{CCSXML}

\ccsdesc[500]{Computer systems organization~Quantum computing}

\keywords{Quantum Computing, Variational Quantum Algorithms, VQE, QAOA}

\maketitle

\pagestyle{plain}

\vspace{-1mm}

\section{Introduction}
\label{sec:introduction}

\rev{Quantum computing promises to tackle problems that are intractable for classical machines by exploiting uniquely quantum phenomena such as superposition and entanglement~\cite{preskill2018quantum,ludmir2024pachinqo}. However, today’s quantum devices and their building blocks (i.e., qubits) suffer from a variety of hardware noise effects such as short coherence times, limited connectivity, and imperfect gate operations~\cite{javadi2024quantum}. These limitations manifest as output error when a piece of quantum code is run on \textbf{high-noise (i.e., low-fidelity)} qubits, resulting in \textbf{low output fidelity}. As a consequence, reliably executing deep or long-running quantum algorithms remains a significant challenge. Much of current research, therefore, focuses on noise-tolerant, hybrid quantum-classical approaches that can extract useful results from imperfect hardware~\cite{ravi2022vaqem,galda2021transferability}.}

Variational quantum algorithms (VQAs) are among the most promising candidates for achieving quantum utility on near-term quantum computers~\cite{preskill2018quantum,bharti2022noisy,cerezo2021variational}. By combining parameterized quantum circuits (i.e., quantum code) with classical optimization, VQAs offer a noise-tolerant hybrid approach to solving problems in quantum chemistry, optimization, and machine learning (Fig.~\ref{fig:vqe})~\cite{jin2024tetris,havlivcek2019supervised}. However, despite their algorithmic resilience, VQAs remain costly to execute on real quantum hardware. \rev{They require repeated circuit evaluations, often hundreds of iterations per run during the optimization procedure, and are sensitive to device noise, compilation strategy, and hardware connectivity layout~\cite{dangwal2023varsaw,10764550}.}

To mitigate these challenges, scientists and practitioners often compile VQAs to run on the subset of qubits with the highest fidelity on technologies with non-uniform qubit fidelity profiles like superconducting qubit architectures~\cite{dangwal2023varsaw}. While this strategy helps maximize performance (the ability to reach close to the optimal value), it leads to underutilization of the rest of the computer, longer job queues, and lower overall system throughput~\cite{ravi2021quantum,10764550}. Worse still, it assumes a static view of the circuit-to-qubit mapping throughout the VQA execution despite the fact that noise resilience varies significantly across different stages of the optimization~\cite{10764550,ravi2023navigating,10.1145/3622781.3674178}.

Recent work in this area has begun to challenge this static model. For instance, Qoncord~\cite{10764550}, proposes executing VQAs in two phases: an exploratory phase on a low-fidelity machine and a fine-tuning phase on a high-fidelity machine. This coarse-grained scheduling approach demonstrates that different fidelity levels may be appropriate at different stages of execution. However, Qoncord operates at the granularity of entire devices and fails to take advantage of the heterogeneous fidelity landscape within a single quantum computer due to the variety in qubit noise models. Moreover, its binary phase split -- low and then high -- does not capture the subtler fidelity requirements of different optimization paths or ansatzes (circuit structures).

\vspace{2mm}

\noindent\textbf{\sol{}:} Our work takes the idea of fidelity-aware execution further to improve performance, convergence, and system throughput simultaneously. We introduce \sol{}\footnote{\rev{\sol{} is an acronym for ``\underline{N}on-uniform \underline{E}xecution with \underline{S}elective \underline{T}ransitions'', referring to NEST's innovative design of evolving the circuit map during VQA execution.}}\footnote{\sol{} is published in the Proceedings of the ACM SIGMETRICS International Conference on Measurement and Modeling of Computer Systems (SIGMETRICS), 2026.}, a technique that dynamically varies the quantum circuit mapping over the course of VQA execution by leveraging the spatial non-uniformity of quantum hardware noise profiles. Unlike prior approaches that fix the qubit map or switch between a small number of predefined configurations, \sol{} adapts the qubit assignment progressively, improving the fidelity of the mapping across iterations using a fidelity metric called Estimated Success Probability (ESP)~\cite{stein2022eqc,ludmir2024parallax,ludmir2024pachinqo,brandhofer2023optimal,tannu2019ensemble,xie2021mitigating}.

To ensure that these transitions do not introduce instability into the optimization process, \sol{} introduces a structured \textit{qubit walk} -- a methodical and incremental remapping of individual qubits. \rev{This opportunity is afforded due to the heterogeneous noise profile of qubits on the same chip on superconducting architectures, which is not afforded on architectures with homogeneous qubits.} This gradual adjustment avoids sharp discontinuities in the optimization landscape, which could arise from abrupt map switches, and allows the optimizer to adapt smoothly. In addition to improving performance and convergence, \sol{} enables a second optimization dimension: concurrency. By assigning non-overlapping sets of qubits with appropriate fidelity to different VQAs, \sol{} supports the co-location of multiple jobs on the same quantum processor. This multi-programming capability significantly improves system throughput while maintaining the performance and convergence behavior of each individual job. \textit{Together, these ideas allow \sol{} to address three core challenges in near-term quantum computing: how to (1) improve VQA performance (i.e., proximity to the optimal value) in the presence of hardware noise, (2) accelerate the convergence of the optimization process, and (3) increase system throughput through more effective resource utilization.}

\vspace{2mm}

\noindent\textbf{The contributions of our work include:}

\begin{itemize}[leftmargin=*]

    \item We introduce \textbf{\sol{}}, a fidelity-aware execution framework that adapts quantum circuit mapping over the course of VQA execution by leveraging qubit-level fidelity heterogeneity within a quantum computer, as opposed to across computers. 

    \vspace{2mm}
    
    \item Inspired by classical machine learning precision results~\cite{10.1007/s10994-023-06480-0}, \sol{} explores \textbf{multiple ESP schedules} to demonstrate that varying the fidelity over the course of VQA execution helps the algorithm explore the optimization landscape in a more effective manner than always running on a high-fidelity circuit map.

    \vspace{2mm}

    \item We design and implement a \textbf{qubit walk} strategy that incrementally transitions circuit mappings to improve stability and convergence, avoiding sharp disruptions in the cost landscape.

    \vspace{2mm}

    \item We demonstrate how \sol{} enables \textbf{multi-programming} of multiple VQAs on a single machine, assigning disjoint qubit subsets and improving system throughput without sacrificing the performance and convergence of VQA algorithms.

    \vspace{2mm}

    \item We implement \sol{} using simulations and real-hardware executions on IBM’s superconducting quantum processors and evaluate it on three molecular Variational Quantum Eigensolver (VQE)~\cite{kandala2017hardware} benchmarks, showing that \sol{} outperforms existing approaches -- BestMap (this technique always executes on the highest-fidelity map~\cite{liu2021relaxed,jin2024tetris,li2022paulihedral,li2024qutracer,wang2022quantumnas}) and Qoncord~\cite{10764550} -- in \textbf{performance}, \textbf{convergence}, and \textbf{system throughput}. For example, on average, \sol{} converges 12.7\% faster than BestMap and 47.1\% faster than Qoncord. 

    \vspace{2mm}

    \item We introduce a realistic, fidelity-weighted \textbf{user cost model} for quantum cloud execution and show that \sol{} reduces user cost by utilizing high-fidelity resources more selectively and efficiently. Compared to \sol{}, on average, users incur a 1.1$\times{}$ higher cost with BestMap and a 2.0$\times{}$ higher cost with Qoncord.

    \vspace{2mm}

    \item \rev{\textbf{\sol{}'s code and dataset are available at:} \textit{\url{https://github.com/positivetechnologylab/NEST}}.}
    
\end{itemize}
\section{Brief and Relevant Background}
\label{sec:background}

\subsection{\rev{Background on Quantum Computing Basics}}

\rev{Quantum computing is built on the principles of quantum mechanics, where information is stored in \textbf{qubits} rather than classical bits. Unlike classical bits, which exist in states $0$ or $1$, qubits can exist in \textbf{superpositions}: $\alpha \ket{0} + \beta \ket{1}$, where $\alpha$ and $\beta$ are complex amplitudes satisfying $|\alpha|^2 + |\beta|^2 = 1$. Qubits can also be \textbf{entangled}, meaning the state of one qubit cannot be described independently of another. Computation is performed by applying a sequence of quantum gates: unitary operations such as single-qubit rotations and two-qubit entangling gates that transform the qubit state. A unitary gate is defined by a matrix $U \in \mathcal{C}^{2^n\times{}2^n}$, such that $U^\dagger U = U U^\dagger = I$, where $I \in \mathcal{C}^{2^n\times{}2^n}$ is the identity matrix and $U^\dagger \in \mathcal{C}^{2^n\times{}2^n}$ is the complex conjugate transpose (adjoint) of $U$. For a unitary matrix, $U^{-1} = U^\dagger$. At the end of execution, qubits are measured, collapsing their superposition into classical outcomes with probabilities dictated by their amplitudes~\cite{han2025enqode}.}

\rev{The current generation of quantum devices, often referred to as Noisy Intermediate-Scale Quantum (NISQ) computers, is limited in qubit number and fidelity. Errors arise from decoherence (finite qubit lifetimes), gate imperfections, and readout noise~\cite{huo2025revisiting,ludmir2025quorum}. These limitations make it difficult to directly run deep quantum circuits or error-corrected algorithms. As a result, much of today’s focus is on hybrid quantum-classical approaches that tolerate noise while exploiting quantum resources.}

\rev{VQAs fall into this category~\cite{cerezo2021variational}. They leverage a parameterized quantum circuit whose parameters are optimized by a classical optimizer to minimize a cost function. By iteratively running shallow quantum circuits and feeding results to a classical feedback loop, VQAs can solve meaningful problems while accommodating hardware noise. We next provide a detailed overview of VQAs before describing our proposed framework, NEST.}

\subsection{Variational Quantum Algorithms}

\begin{wrapfigure}{r}{0.65\linewidth}
    \vspace{-4mm}
    \centering
    \includegraphics[width=0.99\linewidth]{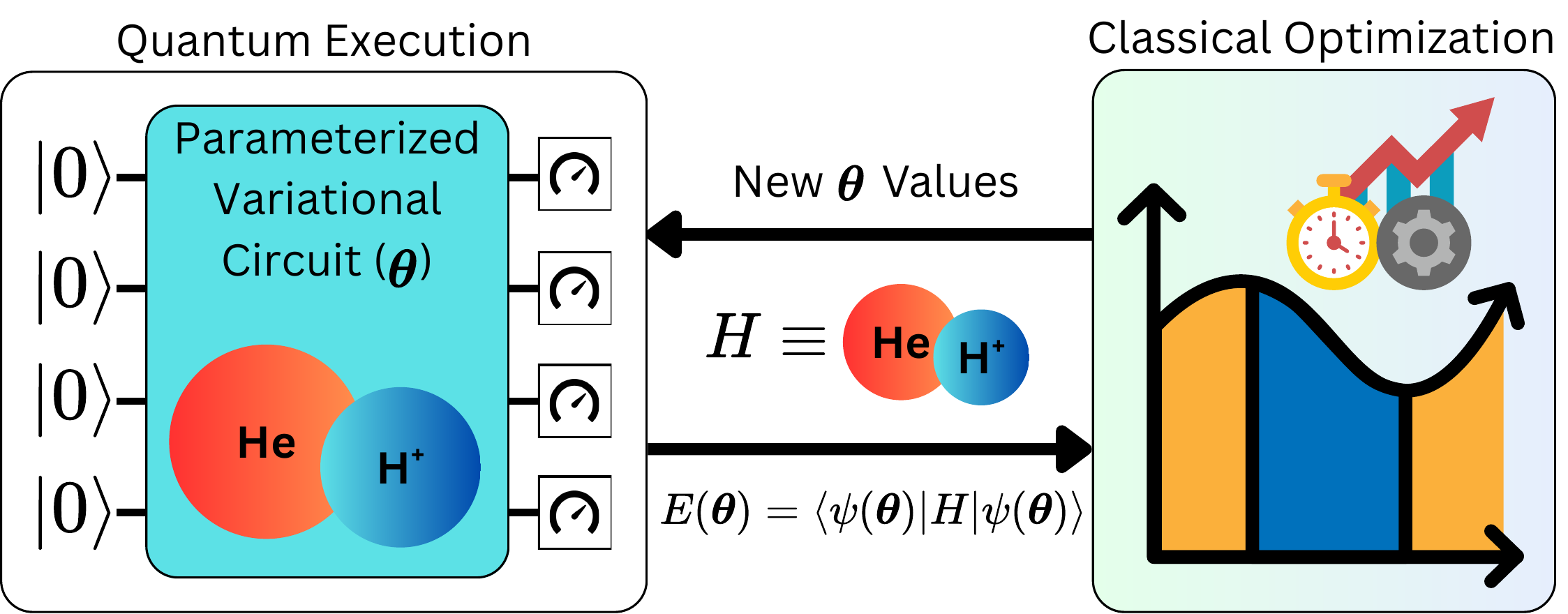}
    \vspace{-3mm}
    \caption{The hybrid quantum-classical execution workflow of a variational quantum algorithm (e.g., VQE~\cite{kandala2017hardware}). \rev{The quantum circuit is shown here with four qubits (horizontal lines). The circuit starts in the ground state, indicated by $\ket{0}$, then the VQA circuit is run, and once completed, the qubits are measured, shown by the meters at the end.}}
    \label{fig:vqe}
    \vspace{-3mm}
\end{wrapfigure}

As shown in Fig.~\ref{fig:vqe}, VQAs are hybrid quantum-classical procedures designed to be resilient to certain types of noise. They parameterize a quantum circuit $U(\boldsymbol{\theta})$ and train its parameters $\boldsymbol{\theta}$ to minimize a cost function $C(\boldsymbol{\theta})$, typically the expectation value of a Hamiltonian with respect to the state produced by the circuit~\cite{cerezo2021variational}.

At a high level, the VQA optimization loop consists of:

\begin{enumerate}
    \item Preparing the quantum state $|\psi(\boldsymbol{\theta})\rangle = U(\boldsymbol{\theta})|0\rangle^{\otimes n}$.
    \item Measuring an observable $H$ to compute the expectation cost value: $C(\boldsymbol{\theta}) = \langle \psi(\boldsymbol{\theta}) | H | \psi(\boldsymbol{\theta}) \rangle$.
    \item Using a classical optimizer to update $\boldsymbol{\theta}$ based on $C(\boldsymbol{\theta})$.
\end{enumerate}

This process is repeated until convergence or a stopping criterion is met. Due to the stochastic nature of quantum measurement, each estimate of $C(\boldsymbol{\theta})$ requires averaging over many circuit executions and measurements (i.e., shots)~\cite{bharti2022noisy}.

\vspace{2mm}

\noindent\textbf{Variational Quantum Eigensolver (VQE).} VQE is a prominent example of a VQA, originally proposed for estimating the ground state energy of a molecular Hamiltonian. The Hamiltonian is expressed as a weighted sum of Pauli terms: $H = \sum_{j} c_j P_j$, where $P_j \in \{I, X, Y, Z\}^{\otimes n}$ are $n$-qubit Pauli strings, and $c_j$ are real coefficients derived from a basis transformation of the molecular problem~\cite{cerezo2021variational}. To estimate the energy $E(\boldsymbol{\theta}) = \langle \psi(\boldsymbol{\theta}) | H | \psi(\boldsymbol{\theta}) \rangle$, each term $P_j$ is measured independently. Thus, the total number of measurements scales with the number of non-commuting $P_j$s~\cite{verteletskyi2020measurement}.

\subsection{Quantum Computing in the Near Term}

The quantum computers available today, including those accessible through cloud providers such as IBM and AWS Braket, operate under severe hardware constraints~\cite{preskill2018quantum}. Most commercial systems are based on superconducting transmon qubits, which are physically realized as anharmonic oscillators and controlled using microwave pulses~\cite{krantz2019quantum}. Despite recent advances in device engineering, these machines exhibit short coherence times, limited qubit connectivity, and gate operations with non-negligible error rates~\cite{bharti2022noisy}.

\rev{On superconducting systems, each qubit is subject to two primary sources of decoherence: energy relaxation (characterized by $\overline{T_1}$) and dephasing (characterized by $\overline{T_2}$)~\cite{blais2021circuit}.} These lead to stochastic and coherent noise processes that affect quantum state evolution. Gate fidelities for single-qubit operations are generally above 99.9\%, but two-qubit gate fidelities often range between 97\% and 99.5\%, with error rates that vary significantly between devices and qubits~\cite{arute2019quantum}. Measurement error further compounds the noise, with typical readout fidelities in the range of 95\%–99\%~\cite{kandala2017hardware}.

The physical layout of superconducting systems also imposes architectural constraints. Devices like IBM’s 27-qubit Falcon or 127-qubit Eagle processors employ fixed topologies, where two-qubit gates are only available between specific pairs of qubits -- IBM uses the heavy-hex qubit connectivity topology shown in Fig.~\ref{fig:vqa_run}~\cite{jin2024optimizing}. This leads to additional SWAP operations during circuit transpilation to help distant qubits interact, increasing circuit depth and noise exposure~\cite{chamberland2020topological}. Consequently, compilation decisions, especially the choice of which qubits to map a circuit to, directly affect both fidelity and runtime\cite{ji2025algorithm}. We also note that circuit mapping will continue to be important even beyond the near term and into the early Fault-Tolerant Quantum Computing (FTQC) era because even on computers with error correction, it is desirable to map to high-fidelity qubits to execute in the error correction regime and to reduce the error correction overhead~\cite{benito2025comparative}. 

\subsection{The Impact of Hardware Noise on VQAs}

Quantum hardware noise manifests in two main ways when a quantum circuit is executed:

\begin{itemize}[leftmargin=*]
    
    \item \textbf{Coherent errors}: Structured, repeatable errors such as calibration drift or crosstalk, which distort the intended gate operations in a biased way. These can be particularly damaging over many circuit iterations when new parameter values rely on prior parameter values during the optimization procedure~\cite{murali2020software}.
    
    \vspace{2mm}

    \item \textbf{Stochastic errors}: Random bit flips and phase flips due to interaction with the environment. These errors accumulate with circuit depth (length of the quantum code) and are modeled by depolarizing, damping, and dephasing channels~\cite{huang2021logical}.
    
\end{itemize}

\begin{wrapfigure}{r}{0.58\textwidth}
    \centering
    \includegraphics[width=0.98\linewidth]{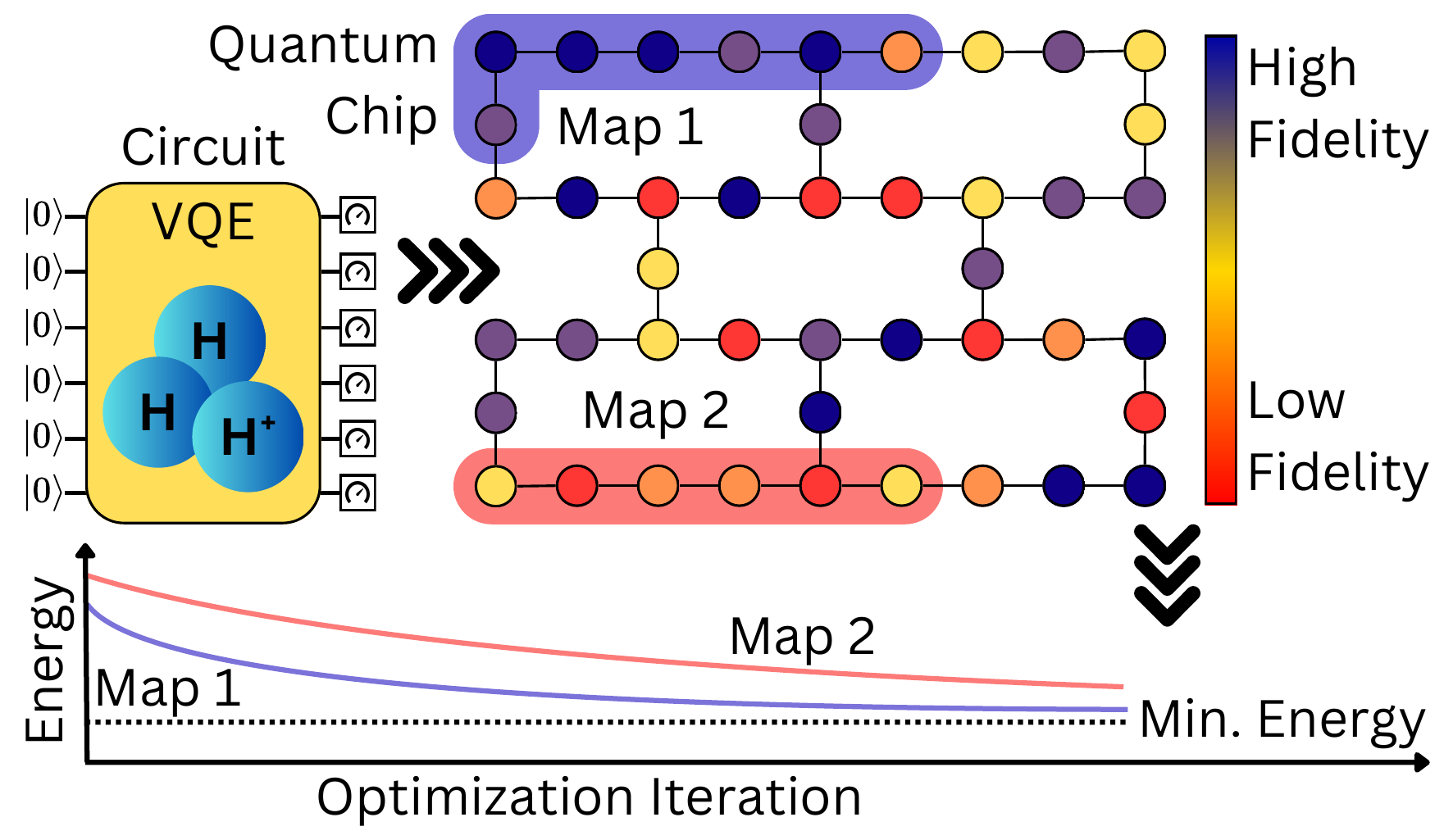}
    \vspace{-4mm}
    \caption{When a variational quantum eigensolver program is run on a high-fidelity map (Map 1), it achieves a closer to optimal performance (closer to the actual minimum energy), than when it is run on a low-fidelity map (Map 2)~\cite{tannu2019not}. Note: the quantum chip is laid out in IBM's heavy-hex topology~\cite{jin2024optimizing}. In this chip picture, the circles are the qubits, and the lines connecting them reflect qubit connections.}
    \label{fig:vqa_run}
    \vspace{-5mm}
\end{wrapfigure}

The effect of noise is most apparent in iterative algorithms like variational quantum algorithms, where the quantum circuit is repeatedly executed with updated parameters~\cite{10764550}. Small amounts of gate or readout error can quickly compound, degrading both the optimization signal and the quality of the final output. In near-term systems, this noise is highly heterogeneous -- certain qubits and gates exhibit lower error rates than others~\cite{tannu2019not}.

\rev{As a result, as shown in Fig.~\ref{fig:vqa_run}, conventional wisdom has been that compilation choices that favor high-fidelity circuit maps within the chip can lead to better performance (closer to the optimal cost or minimum energy in the case of VQE)~\cite{murali2019noise}. To determine the high-fidelity regions, prior work has used the Estimate Success Probability (ESP) to quantitatively assess quantum circuit reliability by integrating gate errors and decoherence effects~\cite{stein2022eqc,ludmir2024parallax,ludmir2024pachinqo,brandhofer2023optimal,tannu2019ensemble,xie2021mitigating}. ESP is computed as:
\begin{equation}
\text{ESP} = \left( \prod_{i=1}^{n} P_{\text{success}}(g_i) \right) \cdot e^{-\frac{d \cdot t_g}{\overline{T_1}}} \cdot e^{-\frac{d \cdot t_g}{\overline{T_2}}}
\label{eq:esp}
\end{equation}
Here, $P_{\text{success}}(g_i)$ represents individual gate success probabilities (fidelities) of gate $g_i$ within the circuit (there are $n$ gates in total in the circuit), and the exponential terms account for decoherence based on circuit depth $d$, the average gate execution time $t_g$, and coherence times $\overline{T_1}$ and $\overline{T_2}$. Higher ESP values indicate more reliable execution. ESP is a helpful metric for estimating what the output fidelity (but not the output itself) of a circuit will be when run on a given computer region without actually running the circuit (which is a high-overhead procedure). We use the ESP metric for our work as it can be computed prior to circuit execution to aid compilation. }

\subsection{\rev{Terminology and Notation}}

\rev{For clarity, we summarize below the key terms used throughout the paper:}

\begin{itemize}[leftmargin=*]
    %\item \rev{\textbf{NEST.} NEST stands for \emph{Non-uniform Execution with Selective Transitions}. It is our proposed framework that dynamically remaps quantum circuits during VQA execution by progressively improving the fidelity of qubit assignments.}

    %\vspace{2mm}
    
    \item \rev{\textbf{Iteration.} An iteration refers to a single update step in the VQA optimization loop. Each iteration consists of preparing the parameterized quantum circuit, executing it on hardware or simulation, measuring the cost function, and updating the circuit parameters via a classical optimizer.}

    \vspace{2mm}
    
    \item \rev{\textbf{Cycle.} A cycle is a higher-level grouping of multiple iterations. Within a cycle, the same qubit-to-circuit mapping is maintained. NEST transitions between different circuit mappings across cycles in order to improve stability and reduce remapping overhead, as described in Sec.~\ref{sec:design}.}

    \vspace{2mm}

    \item \rev{\textbf{Fidelity.} Fidelity quantifies the reliability of a quantum component or computation. Higher fidelity indicates more accurate execution and lower susceptibility to noise.}
    \begin{itemize}
        \item \rev{\emph{Qubit/region fidelity} refers to the success probability of executing a gate on a given qubit/region, e.g., the probability that a single-qubit or two-qubit operation performs as intended.}
        \item \rev{\emph{Circuit fidelity} (also expressed through metrics such as ESP) reflects the overall reliability of an entire circuit execution, obtained by combining gate fidelities with decoherence factors ($\overline{T_1}$, $\overline{T_2}$) and circuit depth.}
    \end{itemize}

\end{itemize}

\rev{Next, we discuss the motivation and timely reason for this work.}

\section{Motivation for \sol{}}
\label{sec:motivation}

VQA’s reliance on repeated quantum circuit execution makes them particularly sensitive to hardware noise. Even small amounts of gate, readout, or decoherence error can compound across iterations, ultimately leading to convergence failures or suboptimal solutions~\cite{wang2021noise,tannu2019not}. While running VQAs on high-fidelity qubit subsets can mitigate this issue, it leads to poor system throughput and underutilization of low-fidelity qubits~\cite{10764550,tannu2019not}. This trade-off between fidelity and efficiency has motivated recent work to explore more flexible execution strategies.

\begin{wrapfigure}{r}{0.65\linewidth}
    \centering
    \includegraphics[width=0.99\linewidth]{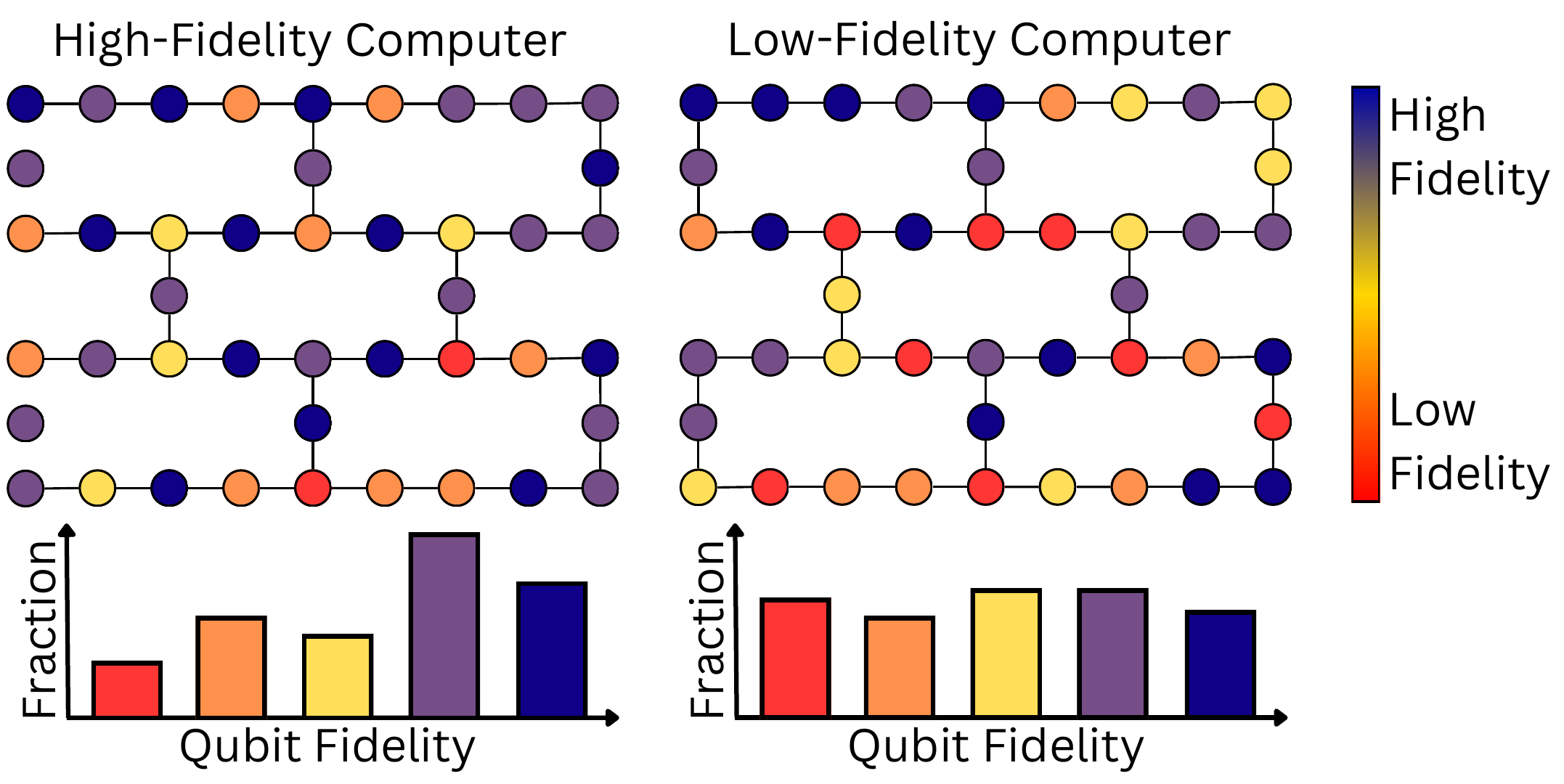}
    \vspace{-3mm}
    \caption{Qoncord~\cite{10764550} models individual computers as high-fidelity devices (the ones with more high-fidelity qubits) and low-fidelity devices (the ones with fewer high-fidelity qubits); however, an opportunity exists within a computer to generate high-fidelity and low-fidelity maps by exploiting qubit noise profile heterogeneity~\cite{murali2019noise}.}
    \label{fig:qoncord_model}
    \vspace{-2mm}
\end{wrapfigure} 

Qoncord~\cite{10764550} takes an important first step in this direction by proposing a two-phase execution model: run early-stage VQA iterations on a low-fidelity machine to explore the optimization landscape, then switch to a high-fidelity machine for fine-tuning. This model captures a useful insight -- early iterations are more noise-tolerant than later ones -- but suffers from several key limitations. First, Qoncord operates at device granularity. It assumes a quantum cloud with distinct low-fidelity and high-fidelity machines and schedules jobs across them. However, in practice, modern superconducting systems already exhibit significant intra-device fidelity heterogeneity~\cite{stein2023hetarch}. As shown in Fig.~\ref{fig:qoncord_model}, a single chip contains regions of high- and low-fidelity qubits. By not exploiting this internal variation, Qoncord leaves substantial opportunity on the table.

Second, Qoncord adopts a binary phase model -- early iterations are noisy (can be run on a low-fidelity circuit map), and later ones are precise (must be run on a high-fidelity circuit map) -- which oversimplifies the dynamics of the variational optimization. In practice, there is a spectrum of possible circuit maps, from low-fidelity (low-ESP) ones to high-fidelity (high-ESP) ones. Thus, treating this as a binary leaves considerable optimization opportunities untapped. Third, Qoncord’s design constrains each phase to a fixed fidelity level. The initial low-fidelity phase remains noisy even as the optimizer begins to converge, while the final high-fidelity phase runs on expensive hardware even if convergence has plateaued. As a result, Qoncord often pays the cost of high-fidelity execution without fully leveraging it for improved performance outcomes. Lastly, Qoncord utilizes disjoint device allocations for different jobs, running only one job on a computer at a time, limiting opportunities for concurrency and co-location. This leads to lower system throughput, especially when most jobs (including non-VQA jobs) compete for the same high-fidelity devices.

These limitations motivate the need for a more fine-grained, adaptive approach. Rather than viewing fidelity as a binary switch across devices, we argue that fidelity should be treated as a tunable spectrum within a single quantum processor. \sol{} is built around this idea. \textit{By leveraging intra-device qubit fidelity heterogeneity, progressively improving circuit map quality over time, and enabling selective yet stable transitions across mappings, \sol{} provides a flexible and efficient execution strategy for variational algorithms—while supporting multi-programming to further improve system throughput.} The next section describes how we achieve this.
\section{\sol{}'s Design Decisions and Features}
\label{sec:design}

%We now describe the design decisions and features of \sol{}.

\subsection{\sol{}'s Exploration of Different ESP Schedules for Circuit Map Selection}

\begin{wrapfigure}{r}{0.5\linewidth}
    \vspace{-0mm}
    \centering
    \includegraphics[width=0.98\linewidth]{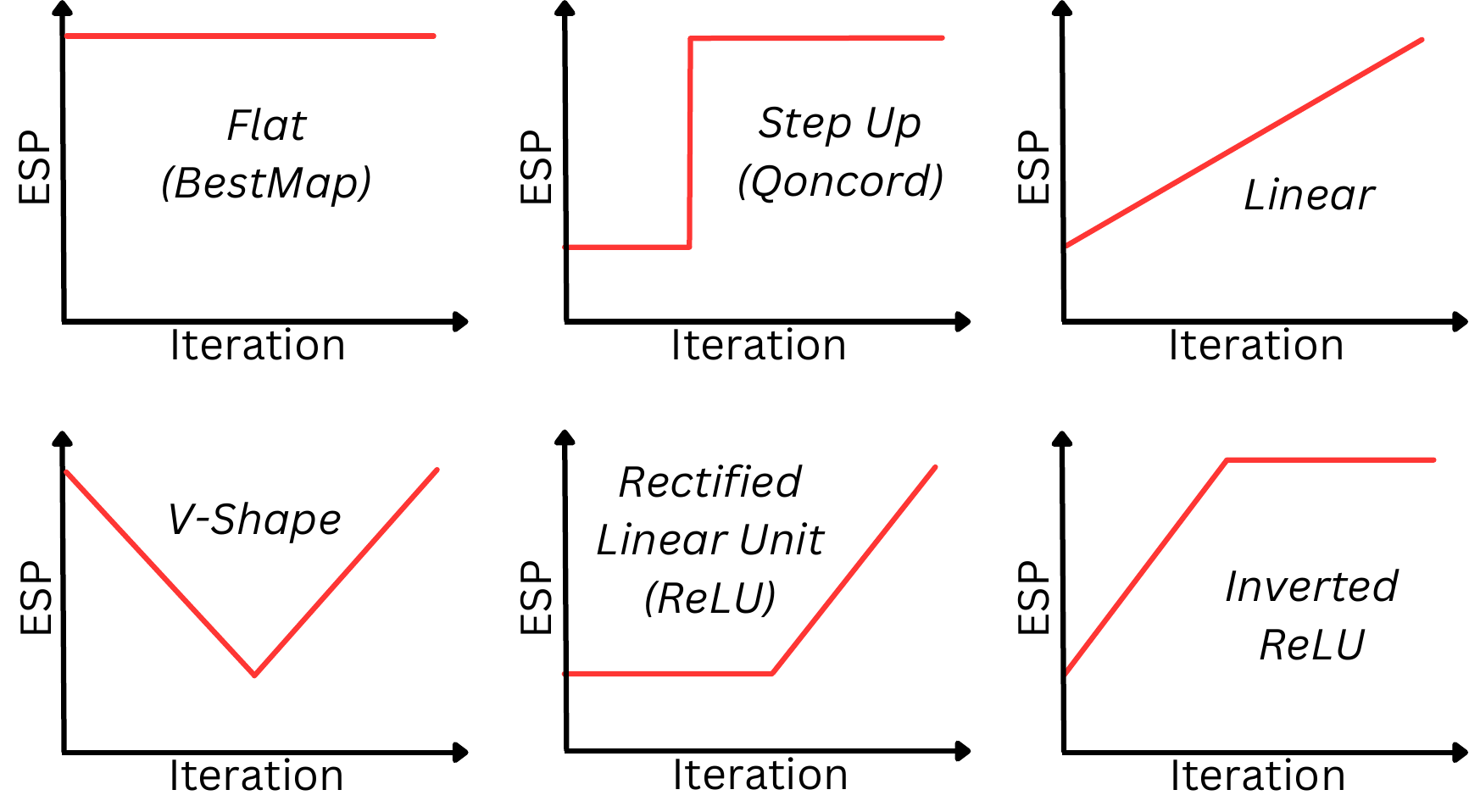}
    \vspace{-3mm}
    \caption{The ESP schedules we explored and how they compare to the optimal map and Qoncord techniques.}
    \label{fig:esp_curves}
    \vspace{-3mm}
\end{wrapfigure}

The first key design decision of \sol{} is to determine how the fidelity of the circuit map should evolve over the course of a VQA execution. Since VQAs are iterative optimization procedures, early iterations benefit from noise-tolerant, exploratory updates, while later iterations demand more precise, high-fidelity execution. To capture this dynamic fidelity requirement, we evaluate six fidelity evolution schedules, each defining a different trajectory for how the ESP of the selected circuit map changes across optimization iterations. These schedules are visualized in Fig.~\ref{fig:esp_curves}.

\vspace{2mm}

%We evaluated six ESP schedules: (1) \textbf{Flat.} The ESP remains constant at a high value throughout execution. This is equivalent to the BestMap strategy, where the highest-fidelity circuit map is selected once and reused for the entire execution \todo{[X]}. (2) \textbf{Step Up.} Execution begins on a low-fidelity map and switches to a high-fidelity map after a fixed point. This is conceptually identical to Qoncord’s two-phase model. (3) \textbf{Linear.} The ESP increases gradually and linearly across all iterations, with the circuit map improving incrementally from low to high fidelity. (4) \textbf{V-Shape.} The ESP first decreases and then increases. This schedule is included to test whether a transient dip in fidelity might encourage broader exploration. (5) \textbf{ReLU.} The ESP stays flat at a low value for the first several iterations and then increases linearly to a high value—motivated by classic curriculum learning strategies. (6) \textbf{Inverted ReLU.} The ESP increases quickly over the first several iterations and then plateaus at a high value for the remainder of the execution. This schedule is designed to enable rapid landscape exploration followed by stable fine-tuning.

\noindent\textbf{The ESP Schedules.} Let $\sigma_t$ denote the ESP at iteration $t$, with $\sigma_{\min}$ and $\sigma_{\max}$ representing the minimum and maximum achievable ESP values for a given circuit. We define $T$ as the total number of iterations in the optimization process. We comprehensively detail the ESP schedules considered in this work below:

\begin{itemize}[leftmargin=*]
    \item \textbf{Flat Schedule.}  This corresponds to using the highest-fidelity circuit map (BestMap~\cite{liu2021relaxed,jin2024tetris,li2022paulihedral,li2024qutracer,wang2022quantumnas}) consistently throughout the execution. The circuit configuration remains unchanged, maintaining a constant ESP value $\sigma_{\max}$ across all iterations.
    \begin{equation*}
    \sigma_t = \sigma_{\max} \forall t \in [0, T]
    \end{equation*}
    
    \vspace{1mm}
    
    \item \textbf{Step Up Schedule.} Execution begins on a low-fidelity map and switches to a high-fidelity map after a fixed point. This is conceptually identical to Qoncord’s two-phase model.
    \begin{equation*}
    \sigma_t = 
    \begin{cases}
    \sigma_{\min}, & \text{if}~~ \frac{t}{T} < \alpha \\
    \sigma_{\max}, & \text{if}~~ \frac{t}{T} \geq \alpha
    \end{cases}
    \end{equation*}
    
    Here, $\alpha \in (0, 1)$ represents the fraction of total iterations at which the transition occurs. We set $\alpha = \frac{1}{2}$ in our experiments after tuning to find the best-performing value. %Following the implementation in Qoncord's two-phase model, we set $\alpha = XXX$ for our experiments, dividing execution into equal exploration and exploitation phases.

    \vspace{2mm}
    
    \item \textbf{Linear Schedule}:  The ESP increases linearly from $\sigma_{\min}$ to $\sigma_{\max}$ over the course of the execution, with improvements to the circuit map at each iteration to achieve the corresponding ESP values.
    \begin{equation*}
    \sigma_t = \sigma_{\min} + \frac{t}{T} \cdot (\sigma_{\max} - \sigma_{\min})
    \end{equation*}
    
    \vspace{1mm}
    
    \item \textbf{V-Shape Schedule}: This schedule begins at $\sigma_{\max}$, decreases linearly to $\sigma_{\min}$ at $t = \frac{T}{2}$, and then increases linearly back to $\sigma_{\max}$ by $t = T$. The temporary reduction in fidelity encourages broader exploration of the solution space before converging toward high-fidelity mappings.
    \begin{equation*}
    \sigma_t = 
    \begin{cases}
    \sigma_{\max} - \frac{2t}{T} \cdot (\sigma_{\max} - \sigma_{\min}), & \text{if}~~ \frac{t}{T} < \frac{1}{2} \\
    \sigma_{\min} + \frac{2(t - T/2)}{T} \cdot (\sigma_{\max} - \sigma_{\min}), & \text{if}~~ \frac{t}{T} \geq \frac{1}{2}
    \end{cases}
    \end{equation*}
    
    \vspace{2mm}
    
    \item \textbf{ReLU Schedule.} The ESP stays flat at a low value for the first several iterations and then increases linearly to a high value -- motivated by classic curriculum learning strategies~\cite{bengio2009curriculum}.
    \begin{equation*}
    \sigma_t = 
    \begin{cases}
    \sigma_{\min}, & \text{if}~~ \frac{t}{T} < \beta \\
    \sigma_{\min} + \frac{t - \beta T}{(1-\beta)T} \cdot (\sigma_{\max} - \sigma_{\min}), & \text{if}~~ \frac{t}{T} \geq \beta
    \end{cases}
    \end{equation*}
    
    Here, $\beta \in (0, 1)$ represents the fraction of iterations maintained at $\sigma_{\min}$ before beginning the linear increase. We set $\beta = \frac{1}{3}$ in our experiments based on the best tuning effort.

    \vspace{2mm}
    
    \item \textbf{Inverted ReLU Schedule.} The inverse of the ReLU schedule.
    \begin{equation*}
    \sigma_t = 
    \begin{cases}
    \sigma_{\min} + \frac{t}{\gamma T} \cdot (\sigma_{\max} - \sigma_{\min}), & \text{if}~~ \frac{t}{T} < \gamma \\
    \sigma_{\max}, & \text{if}~~ \frac{t}{T} \geq \gamma
    \end{cases}
    \end{equation*}
    
    Here, $\gamma \in (0, 1)$ represents the fraction of iterations during which ESP linearly increases. We use $\gamma = \frac{1}{2}$, allowing rapid transition to high-fidelity mappings after initial exploration.
    
    \vspace{2mm}
    
\end{itemize}

Each ESP schedule determines how circuit mappings are selected throughout the optimization process, balancing exploration of the solution space (lower ESP values) with exploitation of promising regions in the optimization landscape (higher ESP values).

\vspace{2mm}

\noindent\textbf{Why Not Use Decreasing ESP Schedules?} We also explored schedules that go from high to low fidelity, but found that these perform poorly in practice. Running on high-fidelity qubits early locks the optimization into narrow basins of the optimization landscape. Subsequent fidelity degradation further corrupts gradients, making recovery difficult. As with classical machine learning, exploration should precede exploitation. Our choice of scheduling is inspired by work in classical precision scheduling for training, which shows that increasing numerical precision over time can improve convergence behavior~\cite{10.1007/s10994-023-06480-0}. \textit{\sol{} leverages the heterogeneity in qubit fidelity as a quantum analog of the controllable precision on classical devices — it controls the circuit map to control the fidelity to improve VQA performance and convergence.}

\vspace{2mm}

\begin{wrapfigure}{r}{0.65\linewidth}
    \centering
    \includegraphics[width=0.99\linewidth]{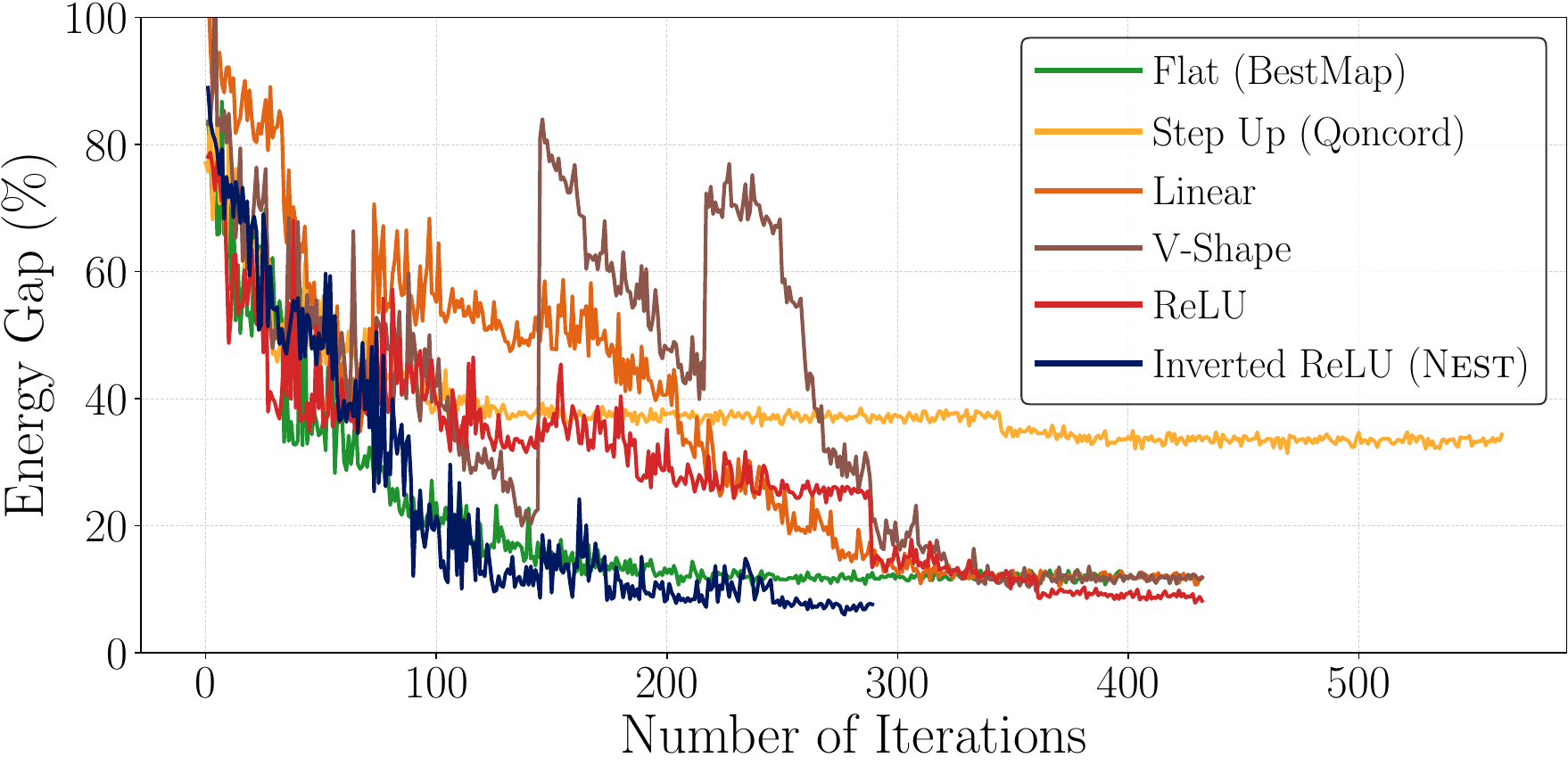}
    \vspace{-3mm}
    \caption{Performance of different ESP schedules on noisy simulations for the H2 molecule VQE circuit~\cite{Utkarsh2023Chemistry,arrazola2021differentiable}; the Inverted ReLU schedule performs the best. The energy gap reflects the percentage difference between the ideal minimum and the achieved minimum cost objective (energy in the case of VQE) -- lower is better. See Sec.~\ref{sec:methodology} for methodological and implementation details for this experiment.}
    \label{fig:esp_perf}
    \vspace{-4mm}
\end{wrapfigure}

\noindent\textbf{Performance Comparison.} Fig.~\ref{fig:esp_perf} shows the energy minimization performance of each of these schedules on a representative VQE benchmark. We find that the Inverted ReLU schedule consistently achieves the best performance. The reasons are intuitive: linearly increasing the ESP early in the run helps the optimizer escape shallow local minima and discover better regions of the landscape. Once the optimization enters a productive region, fixing the fidelity at a high level helps preserve precision during fine-tuning. By contrast, the Flat schedule (BestMap) forgoes any opportunity to adapt fidelity to stage-specific needs and overcommits expensive hardware from the start. The Step Up schedule (Qoncord) suffers from a single hard switch that may occur too early or too late. The Linear schedule transitions too slowly to high-fidelity execution, often wasting early convergence opportunities. The V-Shape schedule temporarily worsens circuit reliability in the middle of optimization, often destabilizing convergence. Finally, the ReLU schedule starts too flat, missing the early gradient signal necessary for the effective exploration of the optimization landscape.

\vspace{2mm}

\begin{figure}
    \centering
    \includegraphics[width=0.8\linewidth]{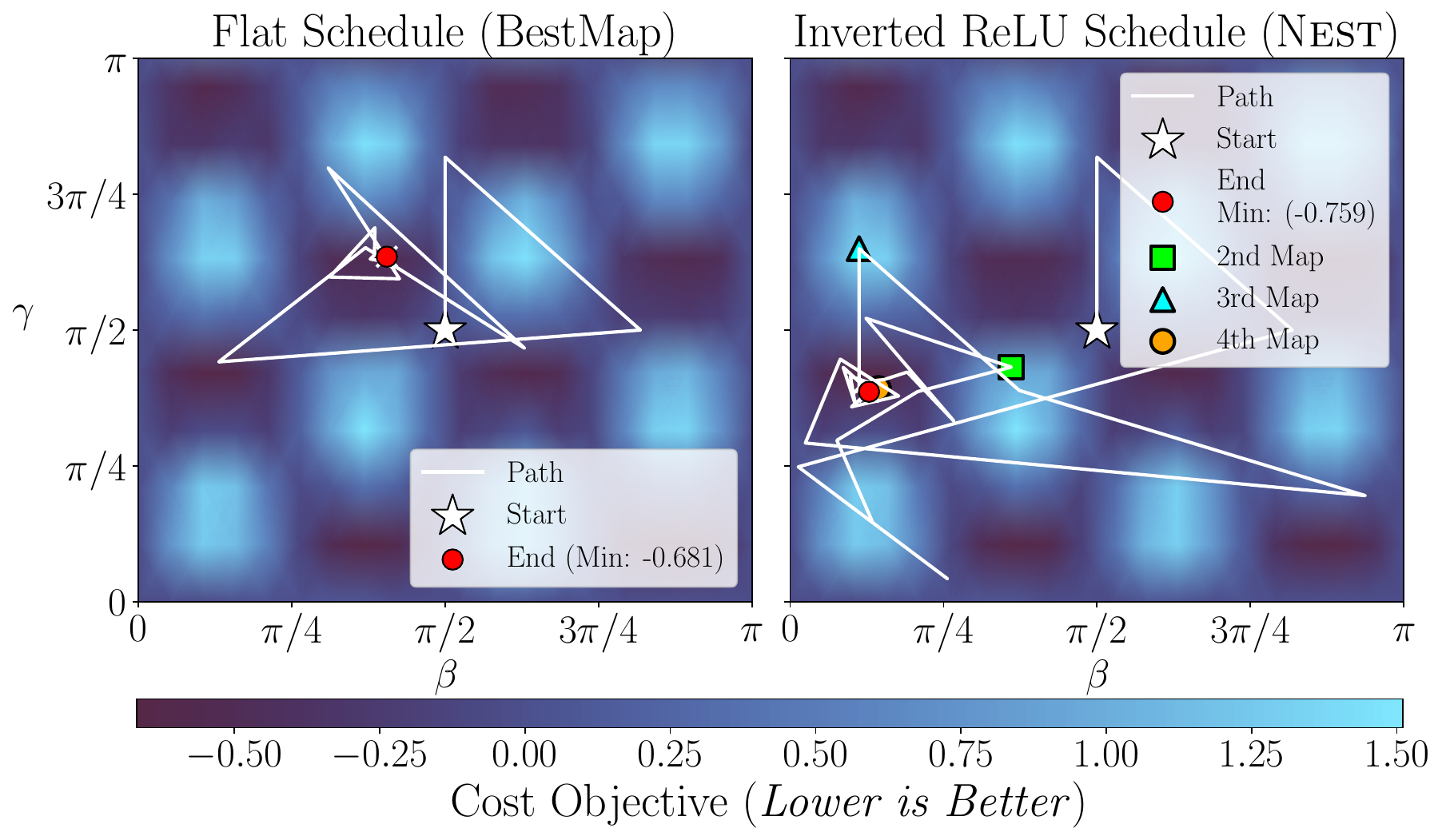}
    \vspace{-3mm}
    \caption{Exploration of the QAOA~\cite{farhi2016quantum} optimization landscape by the (a) Flat (BestMap) and (b) InvertedReLU (\sol) schedules with two parameters: ($\gamma, \beta$). The landscape heatmap is generated using an ideal simulation to get the true values, while the exploration lines are generated using the noise model simulation of the \textit{ibm\_brisbane} quantum computer. Refer to Sec.~\ref{sec:methodology} for further methodological details.}
    \label{fig:qaoa_perf}
    \vspace{-4mm}
\end{figure}

\noindent\textbf{Landscape Exploration with QAOA.} To better understand how ESP schedules impact optimization behavior, we visualize the cost landscape of the variational Quantum Approximation Optimization Algorithm (QAOA)~\cite{farhi2016quantum} circuit as a function of its two variational parameters: $\gamma$ and $\beta$. While this is a very small-scale problem, we select it here as it only has two parameters, which is appropriate for visualization and analysis. We apply a one-layer QAOA to a 10-vertex MaxCut instance with 21 edges, where the objective is to partition the vertices into two sets, maximizing the number of edges crossing the partitions. Fig.~\ref{fig:qaoa_perf} shows a heatmap of the cost function across this parameter space, where darker regions represent lower (better) cost objective values. Compared to the Flat (BestMap) schedule, the path taken by \sol{} using the Inverted ReLU schedule effectively navigates through broader, flatter regions early on, before homing in on precise low-cost basins, thus achieving a lower minimum cost objective than BestMap. This confirms that the variation in the fidelity of qubits across a quantum computer acts as a mechanism for coarse-to-fine search in high-dimensional landscapes.

\vspace{2mm}

Once an ESP schedule is selected -- Inverted ReLU, in our case -- the next question is how to choose concrete circuit maps that match the target ESP value at each stage. The naive approach is to search the entire space of possible maps and select the one closest to the desired ESP at each timestep. However, as we will show next, this strategy introduces large jumps in the circuit layout that harm convergence and inflate search costs. We now describe \sol{}'s qubit walk methodology to mitigate this challenge.

\begin{wrapfigure}{r}{0.65\linewidth}
    \centering
    \vspace{-3mm}
    \subfloat[ESP Discretization]{\includegraphics[width=0.48\linewidth]{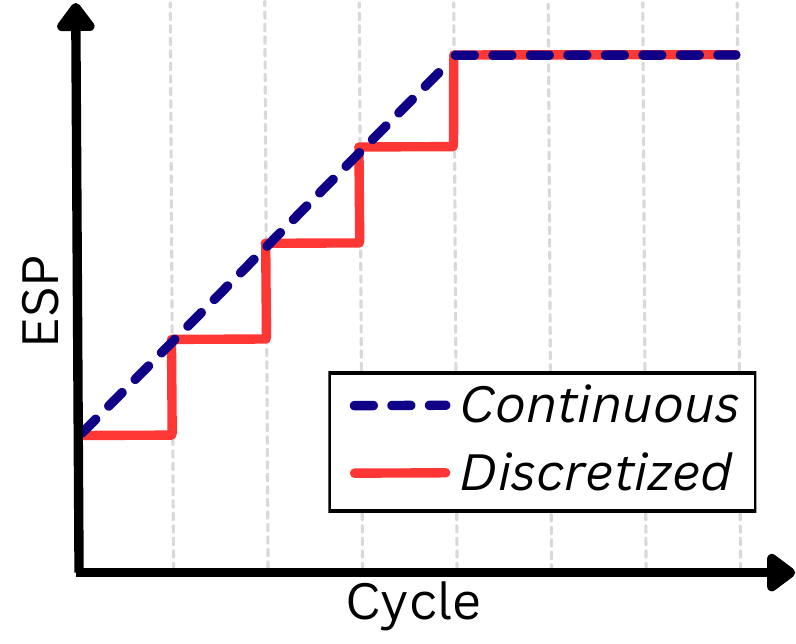}}
    \hfill
    \subfloat[Qubit Jump vs. Qubit Walk]{\includegraphics[width=0.48\linewidth]{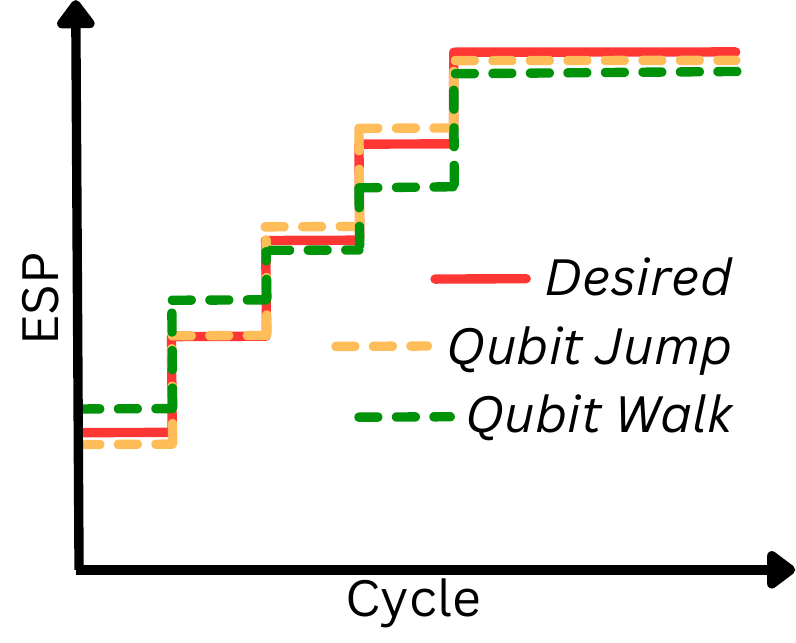}}
    \vspace{-3mm}
    \caption{(a) The ESP schedule has to be discretized to avoid excessive remapping that may hinder convergence stability. (b) Qubit jumps over maps help achieve ESPs that are closer to the desired values, while qubit walks can lead to farther-than-desired ESP values.}
    \label{fig:esp_dev}
    \vspace{-2mm}
\end{wrapfigure}

\subsection{\sol{}'s Qubit Walk Methodology to Reduce Optimization Instability}

As a first step, we must discretize the ESP schedules. Theoretical ESP schedules are continuous, but real quantum systems are discrete: at each iteration, we must choose a concrete circuit map that corresponds to a particular ESP value. Thus, to make the schedule implementable, we must discretize it carefully.

\vspace{2mm}

\noindent\textbf{ESP Schedule Discretization.} \sol{} divides the optimization into multiple \textit{cycles}, where each cycle consists of a fixed number of optimization iterations. Within a given cycle, the circuit map remains constant. This provides two benefits: (1) it reduces the overhead of frequent remapping, and (2) it makes ESP-based scheduling tractable on current hardware. The discretized schedule is illustrated in Fig.~\ref{fig:esp_dev}(a), where the smooth Inverted ReLU curve is approximated in a stepwise fashion at the granularity of a cycle.

\vspace{2mm}

\begin{figure*}
    \centering
    \includegraphics[width=0.99\linewidth]{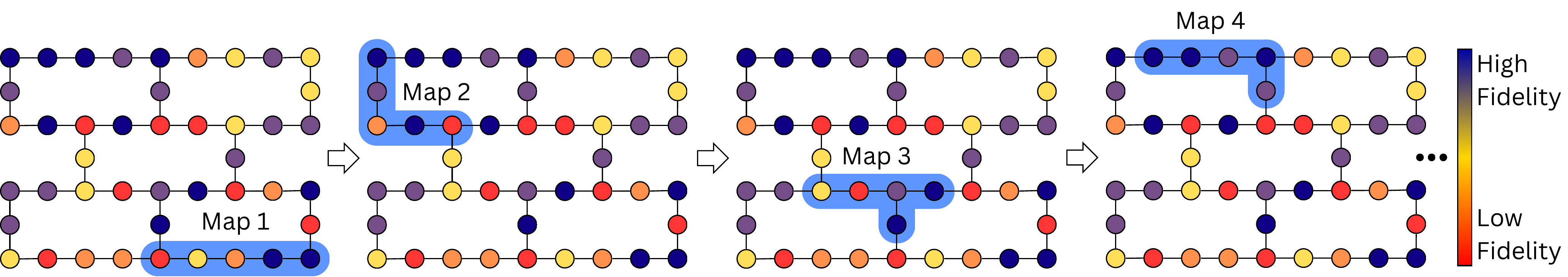}
    \vspace{-1mm}
    \caption{The intuitive qubit jump approach of finding the circuit map closest to the ESP requirement leads to two inefficiencies: (1) It causes unnecessarily large disruptions in the optimization landscape in between the map switches from one cycle to another. (2) As all possible maps have to be explored at each switch to find the best ESP match, this approach is inefficient.}
    \label{fig:jumping}
    \vspace{-4mm}
\end{figure*}

\noindent\textbf{Qubit Jump vs. Qubit Walk.} A straightforward implementation of this discretized schedule would be to, at the start of each cycle, search the space of all possible circuit-to-qubit maps and select the one whose ESP most closely matches the desired value. This approach — referred to as a \textit{qubit jump} — is shown in Fig.~\ref{fig:esp_dev}(b). While this method allows us to match the target ESP values with high fidelity, it suffers from two significant drawbacks. First, as shown in Fig.~\ref{fig:jumping}, the selected maps may be scattered across the chip, causing large spatial dislocations from one cycle to the next. These sharp layout changes destabilize the optimization landscape, leading to erratic convergence. Second, the search space of all circuit maps is combinatorial in the number of qubits, making this approach computationally expensive, especially for larger circuits.

\vspace{2mm}

\begin{figure*}
    \centering
    \includegraphics[width=0.99\linewidth]{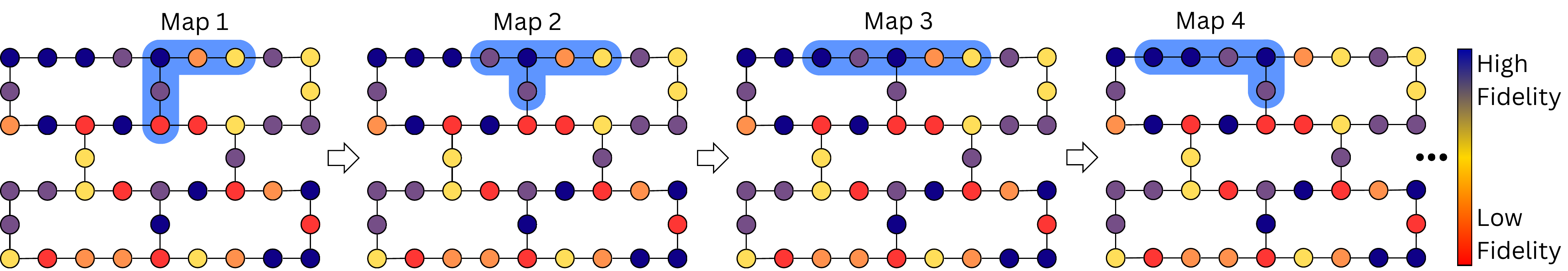}
    \vspace{-1mm}
    \caption{\sol{} carefully navigates the move across circuit maps for a given quantum circuit from the lowest-fidelity map to the highest-fidelity map by walking one qubit at a time to avoid sudden and large disruptions in the optimization landscape. This approach is also more computationally efficient as it reduces the number of eligible maps during the switch.}
    \label{fig:walking}
    \vspace{-4mm}
\end{figure*}

\noindent\textbf{Qubit Walk Methodology.} To mitigate both of these issues, \sol{} introduces a \textit{qubit walk} methodology. Instead of jumping to a new circuit map at each cycle, \sol{} incrementally transitions from one map to the next by walking one qubit at a time. At each transition point, \sol{} greedily considers only the neighboring circuit maps — those that differ from the current map by a small, localized remapping of one or two qubits — and selects the one that brings the ESP closer to the target schedule. As illustrated in Fig.~\ref{fig:walking}, this walk-based approach smooths out circuit transitions, reducing disruptive optimization artifacts and allowing the optimizer to adapt gradually. While qubit walks may not always match the desired ESP as precisely as a global jump (Fig.~\ref{fig:esp_dev}(b)), they strike a better balance between schedule tracking, convergence stability, and runtime overhead. For the same ESP on both methods, however, the walk maintains similar individual single-qubit and two-qubit gate errors at each step, which aids in convergence stability. 

On quantum devices, neighboring qubit sets may also share some hardware-specific noise sources, such as crosstalk, leakage, and other correlated errors, that allow for more similar convergence behavior as opposed to jumps. The walk respects spatial locality, avoids abrupt discontinuities in the cost landscape, and scales more efficiently with larger quantum circuits.

\vspace{2mm}

Through cycle-based discretization and the qubit walk strategy, \sol{} implements a practical and efficient realization of the Inverted ReLU schedule, preserving the benefits of fidelity variation without incurring its potential costs. Beyond stability and efficiency, this walk-based method also unlocks a new capability: since \sol{} only uses a small subset of qubits at each point in time, it opens the door to co-locating multiple jobs on the same quantum computer. We describe this multi-programming extension next.

\subsection{Extending \sol{} to Support Concurrent Runs with Multi-Programming}

An additional benefit of \sol{}'s qubit walk methodology is its natural support for \textit{multi-programming} -- the ability to co-locate multiple VQA jobs on the same quantum processor without significant performance degradation. This is made possible by two key properties of \sol{}: (1) its use of spatially localized circuit maps due to intra-device fidelity heterogeneity, and (2) its gradual, localized map transitions enabled by the qubit walk strategy. Modern superconducting quantum computers exhibit significant variation in qubit fidelity across the chip, with low- and high-fidelity qubits existing throughout the computer. As a result, not all qubits are in use at any given time. This spatial sparsity -- combined with the fact that \sol{} only walks to adjacent maps within a localized region -- naturally enables the scheduler to allocate unused regions of the chip to other VQA jobs. \sol{} exploits this by selecting different fidelity ``zones'' of the device for different VQAs. 

\begin{figure*}
    \centering
    \includegraphics[width=0.99\linewidth]{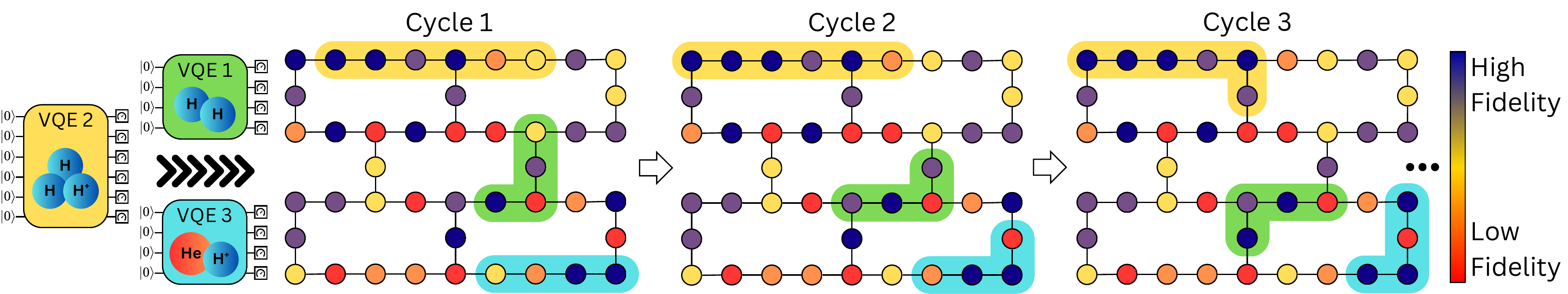}
    \vspace{-2mm}
    \caption{\sol{}'s qubit walk methodology is suitable for and can be extended to co-locate multiple programs (three in this example) on the same quantum chip without impacting the fidelity of the execution to increase the throughput of the system.}
    \label{fig:multi-prog}
    \vspace{-2mm}
\end{figure*}

Fig.~\ref{fig:multi-prog} shows an example of this capability. Here, \sol{} co-locates three VQA jobs on the same quantum processor, each operating on its own subset of qubits. Because each job independently walks through the device in a controlled and non-overlapping way, the fidelity guarantees of each program remain intact. More importantly, this multi-programming does not compromise the performance or convergence of any single job, while significantly improving system throughput. Unlike traditional static circuit mapping approaches, which monopolize the highest-fidelity regions for the entire execution, \sol{} opens up the quantum system to finer-grained resource sharing. In practice, this enables quantum cloud providers to achieve higher utilization without sacrificing the correctness or quality of results, thus increasing system throughput.

With the key design decisions of \sol{} in place, we now summarize its end-to-end execution.

\subsection{Putting Together all the Elements of \sol{}}
\label{subsec: comprehensive_maps}

\begin{figure}
    \centering
    \includegraphics[width=0.8\linewidth]{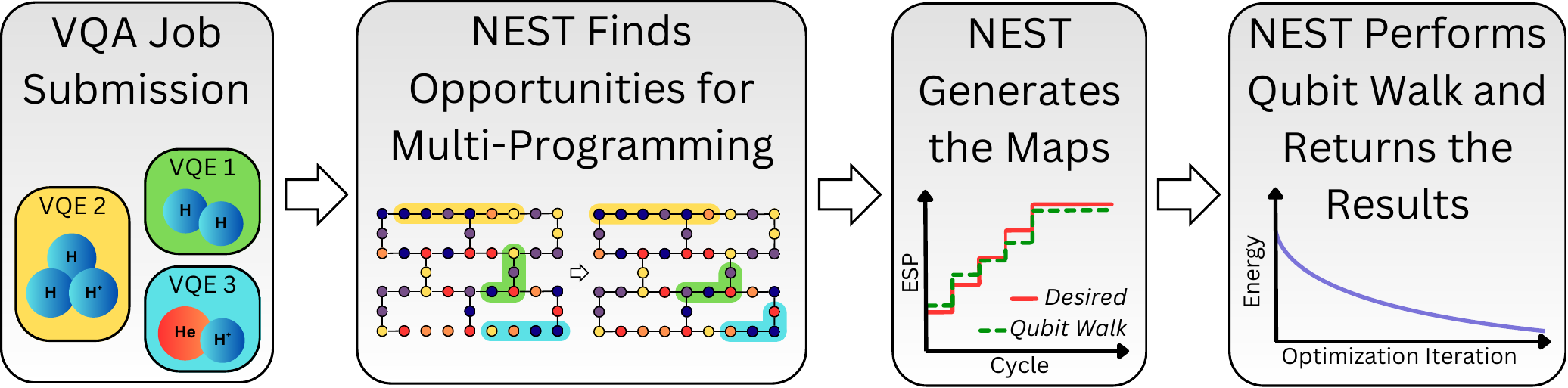}
    \vspace{-2mm}
    \caption{The overall flow of the steps taken by \sol{}.}
    \label{fig:workflow}
    \vspace{-4mm}
\end{figure}

To summarize the design and workflow of \sol{}, we provide the visual shown in Fig.~\ref{fig:workflow}. When a VQA job is submitted to the quantum cloud, \sol{} first finds opportunities for multi-programming to co-locate the job with other concurrently submitted jobs. Once the co-location decisions are made, \sol{} executes the job with the ``Inverted ReLU'' ESP schedules, attempting to find circuit maps that match the schedule as closely as possible while performing its qubit walk routine to help with optimization stability during execution. Finally, the VQA job is run, and the results are returned to the user. The time complexity of running \sol{} for a given VQA is bounded by $O(n(C+Q))$, where  $n$ is the number of qubits in the circuit, $C$ represents the number of cycles in the ESP schedule, and $Q$ is the number of qubits on the quantum computer. The process involves Breadth First Search (BFS) operations on each qubit on the computer, requiring $O(nQ)$ time with heavy-hex connectivity. Then, a circuit map is selected from these to run the first cycle. Then, a qubit walk is performed to find the next map, which requires $O(n)$ operations to explore the neighboring qubits. For $C$ cycles, this process thus has a computational overhead of $O(nC)$, giving an overall time complexity of $O(n(C+Q))$.

Next, we describe the implementation and methodology of \sol{} before evaluating it against competitive techniques.
\section{Implementation and Methodology}
\label{sec:methodology}

\subsection{Experiment Setup}

We implemented circuit construction and noise model simulation using IBM's Qiskit SDK version 1.4.2~\cite{javadi2024quantum}. Noisy circuit simulations were conducted with the Qiskit Aer simulator version 0.17.0. Noise models containing gate errors and coherence times were obtained from five state-of-the-art IBM quantum computers over multiple days: \textit{ibm\_brisbane}, \textit{ibm\_kyiv}, \textit{ibm\_brussels}, \textit{ibm\_sherbrooke}, and \textit{ibm\_strasbourg}~\cite{castelvecchi2017ibm}.

\rev{The noise models are generated by IBM and faithfully emulate the noise characteristics of the real hardware. The noise models contain daily characteristics based on daily benchmarking of qubit properties during calibration~\cite{huo2026anchor,huo2025revisiting}. They contain properties for each qubit, such as $\overline{T_1}$ and $\overline{T_2}$ coherence times, one-qubit gate times, two-qubit gate times, one-qubit error rates, and two-qubit error rates. They also contain information such as the qubit connectivity map. Our real-computer experiments utilize the same computers that we collected the daily noise models for.} However, it is prohibitively expensive and slow (due to long queue times~\cite{ravi2021quantum}) to execute all of our experiments on real hardware, so we perform most of our experiments using noisy simulation runs and use real hardware runs for validation. For optimization, we used the \textit{minimize} function from SciPy with the COBYLA (Constrained Optimization BY Linear Approximation) optimizer~\cite{virtanen2020scipy,powell2007view}.

\subsection{Competitive Techniques}

Experiments using \sol{} were compared with two competitive techniques: BestMap and Qoncord~\cite{10764550}. BestMap is a widely used technique that performs all optimization iterations on a single map that produces the highest ESP~\cite{liu2021relaxed,jin2024tetris,li2022paulihedral,li2024qutracer,wang2022quantumnas}. The optimization procedure for BestMap uses the default step size of 1 and terminates based on the default convergence criteria implemented in the Scipy minimize function. Qoncord initially operates on low-fidelity quantum computers before transitioning to high-fidelity quantum computers. In their approach, Qoncord employs a related metric called the Execution Fidelity Estimator, given by the following:
$$P_{\text{Correct}} = e^{-\frac{CD\frac{\mu_1 G_1 + \mu_1 G_2}{2}}{\overline{T_1}\overline{T_2}}}(1-\gamma)^{G_1}(1-\beta)^{G_2}(1-\omega)^M,$$
which is conceptually the same as the ESP metric (Eq.~\ref{eq:esp})~\cite{stein2022eqc,ludmir2024parallax,ludmir2024pachinqo,brandhofer2023optimal,tannu2019ensemble,xie2021mitigating}, but assumes that all qubits on the computer have the same error rate. The low-fidelity computer implementation uses a step size of 1 and utilizes a tolerance threshold of 0.1 as the termination criterion, whereas the high-fidelity computer implementation operates with a reduced step size of 0.1 and relies on the default Scipy tolerance parameters for termination.

Due to the stochasticity of variational techniques, we run each benchmark with each technique 30 times (e.g., with different seeds and different initializations) and report the mean and the standard deviation across all metrics as appropriate.

\subsection{\rev{VQA Simulation}}

\rev{We briefly outline how VQAs are simulated in our work. The simulation workflow mirrors the hybrid quantum-classical execution model but is implemented entirely within a software environment. The steps are as follows:}

\begin{enumerate}
    \item \rev{\textbf{Circuit construction.} A parameterized quantum circuit $U(\boldsymbol{\theta})$ is generated using Qiskit, with parameters $\boldsymbol{\theta}$ initialized randomly.}
    \item \rev{\textbf{State preparation and execution.} For each iteration, the circuit is executed on the Qiskit Aer simulator under a chosen noise model. The simulator emulates the behavior of IBM superconducting quantum processors, including gate errors, decoherence, and measurement noise. The noise models are generated by IBM based on daily characterization data.}
    \item \rev{\textbf{Measurement and cost evaluation.} The circuit is executed for a fixed number of shots to estimate the expectation value of the cost Hamiltonian $H$. This yields the cost function $C(\boldsymbol{\theta}) = \langle \psi(\boldsymbol{\theta}) | H | \psi(\boldsymbol{\theta}) \rangle$.}
    \item \rev{\textbf{Classical parameter update.} The measured cost is passed to a classical optimizer (we use COBYLA from SciPy), which updates the parameters $\boldsymbol{\theta}$ according to its optimization rule.}
    \item \rev{\textbf{Stopping criterion.} Steps (2)–(4) are repeated until convergence which is as defined for different techniques above.}
\end{enumerate}

\subsection{Resource Availability Protocol}

We simulate the submission of each algorithm 30 times (for each of our evaluated techniques) to ensure statistical robustness. Due to maintenance schedules and queue limitations in quantum computing infrastructure, simultaneous access to all five IBM quantum computers is not guaranteed for real-world scenarios~\cite{ravi2021quantum}. To simulate realistic and reproducible conditions, we set that two of the five computers were available for each experimental run, reflecting typical access constraints in quantum computing applications.

From these two selected computers, the same one was used for both BestMap and \sol{}. For Qoncord, the selection process involved evaluating the ansatz circuit fidelity when transpiled for both available computers, with the lower-fidelity computer designated as the low-fidelity device and the higher-fidelity computer as the high-fidelity device. Note: we do not evaluate different queueing techniques as that is orthogonal to our effort and has already been evaluated in the Qoncord work~\cite{10764550}.

\subsection{Evaluation Benchmarks}

To evaluate our approach, we utilized real-world chemical molecule Hamiltonians to calculate ground-state energies. Three molecules were selected from PennyLane's built-in molecule library~\cite{arrazola2021differentiable,bergholm2018pennylane}: the hydrogen molecule (\htwo{}), the hydrogen molecular ion (\hthree{}), and the helium hydride ion (\heh{})~\cite{Utkarsh2023Chemistry}. Both \htwo{} and \heh{} were represented using 4-qubit Hamiltonians, while \hthree{} required 6 qubits due to its larger molecular structure. For the VQE~\cite{kandala2017hardware} implementation, we employed the EfficientSU2 ansatz~\cite{ravi2022vaqem} from Qiskit with 3 repetitions (\texttt{reps}$=3$)~\cite{javadi2024quantum}.

We also test on a larger circuit, solving the MaxCut problem on 5 real-world graph instances. We used a one-layer QAOA ansatz for training. We employed the same ansatz and transpiler optimizations (all optimizations enabled with optimization level set to 3) for all competitive techniques for a fair comparison. Note that we did not include the simulation of larger molecules, which results in poor output fidelity (almost random outputs generated) due to the quality of the qubits on current systems. Nevertheless, our technique is fundamentally scalable and can be applied to larger algorithms as qubit quality improves.

\subsection{Evaluation Metrics}

We evaluate \sol{} using several different metrics: (1) \textbf{Energy Gap \textit{(lower is better)}.} This metric reflects the ``performance'' of the technique and refers to how close the minimum cost achieved (minimum energy achieved in the case of VQE algorithms) is to the ideal minimum, which is the minimum that would be achieved under ideal simulation conditions (for the VQE algorithms, this is the minimum eigenvalue of the Hamiltonian representing the molecule). We present the metric as normalized percentage distance from the ideal: $\frac{\text{ideal min. energy} - \text{achieved min. energy}}{\text{ideal min. energy}}\times{}100\%$. Lower distance reflects better performance, and higher distance reflects worse performance. (2) \textbf{Number of Iterations \textit{(lower is better)}.} The metric reflects the ``convergence'' of the technique and refers to the number of iterations required for the technique to terminate. It is a proxy for the runtime of the algorithm, as all iterations take the same amount of time. (3) \textbf{System Throughput \textit{(higher is better)}.} This metric simply reflects the number of jobs executed by the quantum cloud service per unit of time. It is inversely proportional to the number of iterations and linearly proportional to the degree of concurrency used to co-locate multiple jobs on the same computer.

\begin{figure*}
    \centering
    \includegraphics[width=0.99\linewidth]{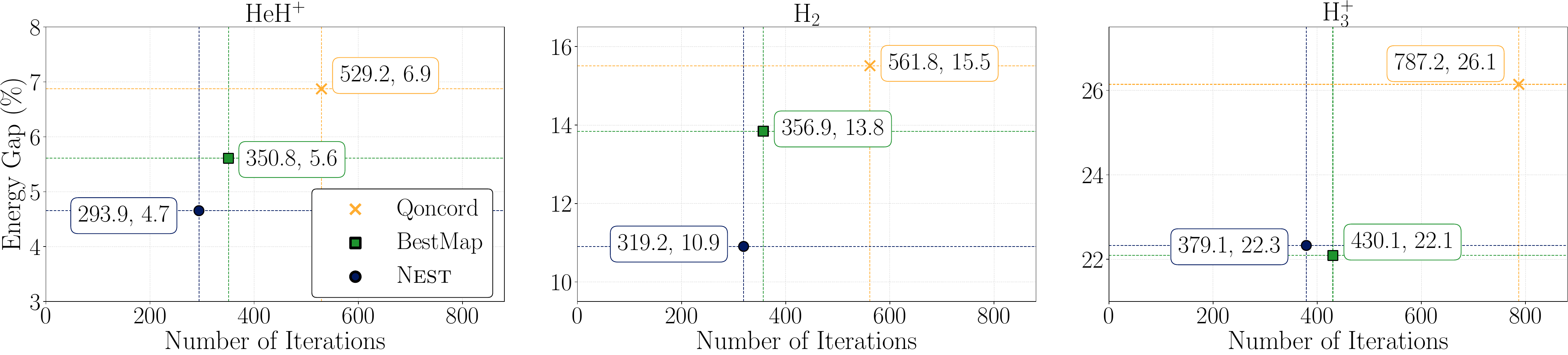}
    \vspace{-3mm}
    \caption{\sol{} achieves a lower energy gap (better performance) than competitive techniques while requiring fewer iterations than them (converging faster) on average across all three VQE algorithms.}
    \label{fig:Eval_2_energy_iterations}
    \vspace{-3mm}
\end{figure*}

\begin{figure*}
    \centering
    \includegraphics[width=0.99\linewidth]{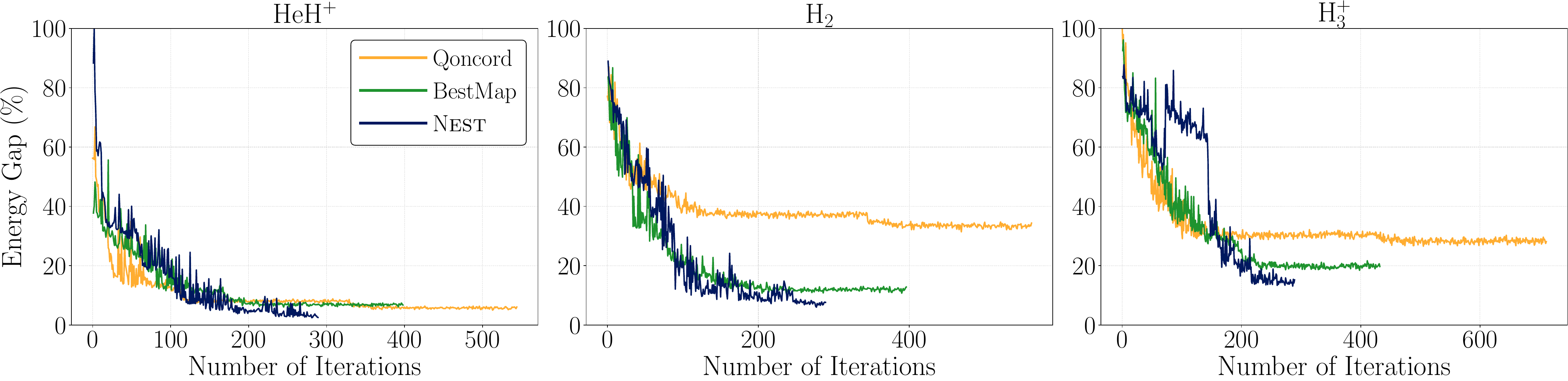}
    \vspace{-3mm}
    \caption{Example energy curves achieved by different techniques during the VQE optimization process for the three algorithms show faster convergence by \sol{} than comparative techniques.}
    \label{fig:Eval_1_energy_curves}
    \vspace{-3mm}
\end{figure*}

(4) \textbf{User Cost \textit{(lower is better)}.} We posit that future quantum computing systems, even early Fault-Tolerant Quantum Computing (FTQC) systems, will charge users based on the fidelity of the qubits that they run on, as the heterogeneous noise profiles of the computers will continue to make the circuit mapping challenge important. This is true even on computers with error correction, where it is desirable to map to high-fidelity qubits to execute in the error correction regime and to reduce the error correction overhead. Thus, this metric should include the average ESP of all the maps used (as maps vary across iterations, we take the average across all maps used for the VQA run), the average circuit depth (which is a proxy for circuit runtime; circuit depth can vary based on the map, so we take the average), and the number of iterations of the VQA (which is a proxy for the entire runtime of the algorithm optimization procedure).

It becomes especially necessary to charge users based on map ESP when running multiple jobs concurrently on the same machine, as the higher-fidelity qubits become prime real estate among the concurrent jobs on technologies with heterogeneous qubit noise profiles. Thus, we propose User Cost $= c\times{}q\times{}\mathbb{E}[\text{ESP}]\times{}\mathbb{E}[d]\times{}I$, where $I$ is the number of iterations, $\mathbb{E}[d]$ is the average circuit depth, $\mathbb{E}[\text{ESP}]$ is the average ESP based on the maps used, $q$ is the number of qubits, and $c$ is a constant applied to calculate the cost in dollar value. With this metric, users are charged based on the length of resource usage, the amount of resources used, and the quality of the resources used. Note: the determination of the exact value of $c$ is dependent on economic and market considerations and is orthogonal to our work, as we use the same $c$ for all techniques.

(5) \textbf{Approximation Ratio \textit{(higher is better)}.} For MaxCut problems, the objective is to identify the maximum cut value for a given graph. The Approximation Ratio is defined as: $\text{Approximation Ratio} = \frac{C_{\text{optimized}}}{C_{\text{ground\_truth}}}$ where $C_{\text{optimized}}$ is the cut value obtained by the algorithm and $C_{\text{ground\_truth}}$ is the optimal maximum cut value for the given graph.
\section{Evaluation and Analysis}
\label{sec:evaluation}

\begin{table}[t]
    \centering
    \caption{\sol{} achieves superior performance on real hardware too, as compared to BestMap and Qoncord.}
    \label{tab:real_machine}
    \vspace{-2mm}
    \renewcommand{\arraystretch}{0.9}
    \begin{tabular}{c>{\columncolor{purple!10}}c>{\columncolor{pink!10}}c>{\columncolor{cyan!10}}c}
        \toprule
        Improvement in \sol{} Energy Gap & \heh{} & \htwo{} & \hthree{} \\
        \midrule
        \makecell{over BestMap Energy Gap (\%)} & 4.3\% & 30.1\% & 13.2\% \\
        \makecell{over Qoncord Energy Gap (\%)} & 15.4\% & 22.0\% & 21.9\% \\
        \bottomrule
    \end{tabular}
    \vspace{-4mm}
\end{table}

\subsection{\sol{} Improves VQA Performance and Convergence over Competitive Techniques}

We begin by presenting the flagship results of \sol{}. Fig.~\ref{fig:Eval_2_energy_iterations} shows the average performance and average convergence for all three techniques across all three molecules. Across the board, \sol{} achieves the lowest energy gap (distance to the ideal minimum energy) and requires the fewest number of iterations. First, we consider the energy gap: one would expect that the BestMap technique should yield the lowest energy gap as the algorithm is run on the same high-ESP circuit map for all of its iterations. However, this is not necessarily the case. \sol{} achieves a lower energy gap than BestMap and, in fact, achieves a far lower energy gap than Qoncord.

For instance, the energy gap for the \htwo{} molecule is 10.9\% with \sol{}, 13.8\% with BestMap, and 15.5\% with Qoncord on average. This is because of \sol{}'s strategy of evolving the circuit maps over the course of the algorithm execution in a manner that carefully improves the ESP of the maps. The switches in the circuit map ESP help \sol{} access different regions of the optimization landscape, which in turn helps \sol{} converge to a better minimum energy than competitive techniques, shortening the energy gap. We also show the energy gap improvement by \sol{} on real hardware in Table~\ref{tab:real_machine}. \sol{} maintains its performance advantage on real hardware over both BestMap and Qoncord for all three VQE molecules.

Further, \sol{} finds a better minimum energy than competitive techniques, while converging much faster than competitive techniques across all molecules. For instance, the number of iterations required for the \heh{} molecule is 294 with \sol{}, 351 with BestMap, and 529 with Qoncord on average. In general, \sol{} converges 12.7\% faster than BestMap and 47.1\% faster than Qoncord. To further analyze how \sol{} achieves this, we present some example energy minimization curves for all three techniques across all three molecules in Fig.~\ref{fig:Eval_1_energy_curves}. The figure highlights some key takeaways: (1) Unlike the other two techniques, \sol{} does not have a monotonically decreasing energy curve; in fact, it has several spikes and bumps in its generally decreasing trend. These spikes happen when \sol{} switches the circuit map from one to another, transitioning from the one with lower ESP to the one with higher ESP. (2) As initially \sol{} runs on low-ESP maps, the general variability in the curve initially is also high compared to competitive techniques.

However, this helps \sol{} explore the optimization landscape in ways that other techniques cannot, thus leading to better performance and a lower energy gap. We also see a slight but noticeable dip in Qoncord when it transitions from its low-ESP map to its high-ESP map. However, this transition happens way too late to have a significant impact or to help explore the optimization landscape beyond the local vicinity. Thus, the transition does not yield much improvement even when given more iterations to run. Further, because Qoncord is only contained to two circuit maps (low-ESP and high-ESP), it does not experience the benefits that \sol{} does by exploring multiple lower-ESP maps in the beginning.

\begin{figure}
    \centering
    \subfloat[Standard Deviation in the Energy Gap]{\includegraphics[width=0.49\linewidth]{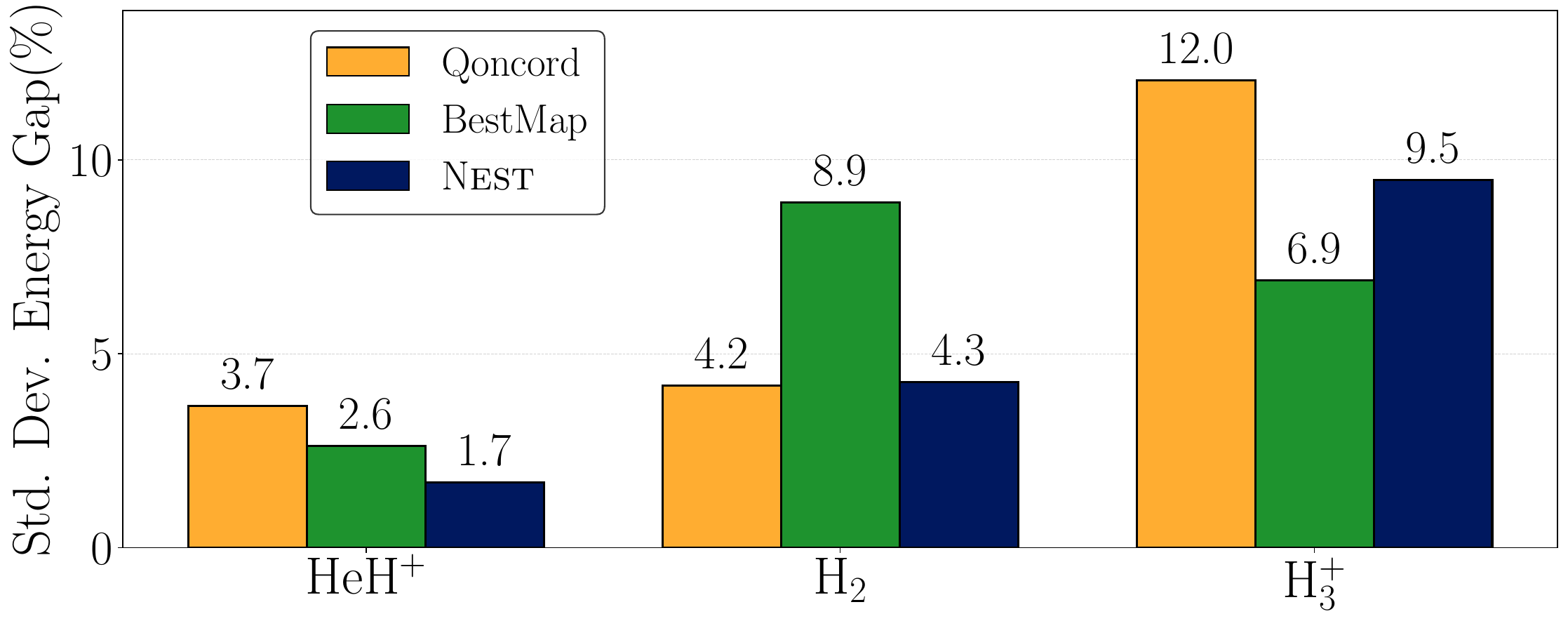}}
    \vspace{-2mm}
    \subfloat[Standard Deviation in the Number of Iterations]{\includegraphics[width=0.49\linewidth]{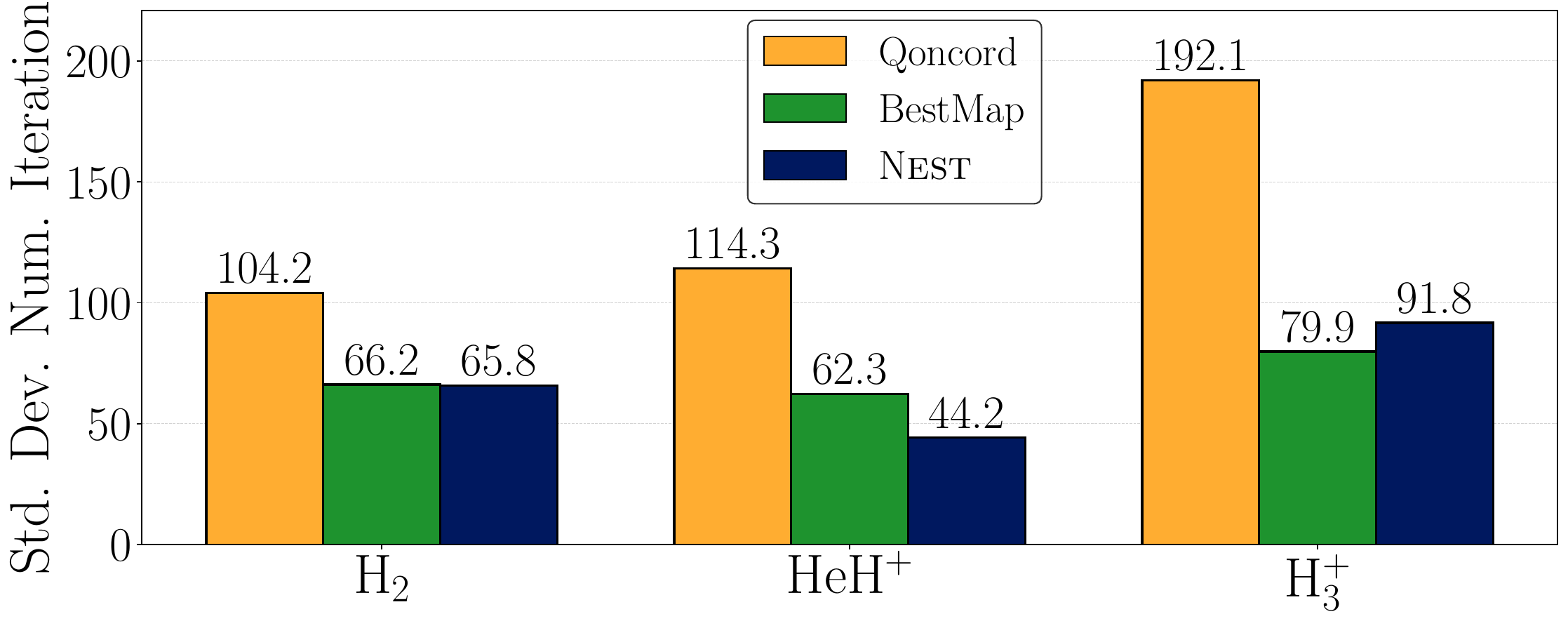}}
    \vspace{-2mm}
    \caption{(a) \sol{} achieves a comparable or lower variability in energy gap as compared to other techniques across all three molecules. (b) \sol{} observes similar trends in terms of variability in the number of iterations. }
    \label{fig:Eval_2_5_stddev_barchart}
    \vspace{-4mm}
\end{figure}

Next, we examine the variability in the runs as we ran VQE on each molecule 30 times (e.g., with different seeds and parameter initialization) with each technique. Fig.~\ref{fig:Eval_2_5_stddev_barchart}(a) shows the variability in the energy gap for all of the techniques. Due to the stochasticity of variational quantum techniques, a certain degree of variability exists with all techniques. In general, \sol{} achieves better or similar variability as other techniques. For the \heh{} molecule, \sol{} achieves the lowest variability, while it ties with Qoncord to achieve the lowest variability for the \htwo{} molecule. For the \hthree{} molecule, Qoncord achieves a considerably high variability; BestMap achieves the lowest variability, but BestMap exhibits a very high variability for the \htwo{} molecule. On average, Qoncord has a 28.9\% higher variability than \sol{} and BestMap has a 19.5\% higher variability than \sol{}. Thus, even though some variability exists across all techniques, \sol{} achieves the most stable performance.

Fig.~\ref{fig:Eval_2_5_stddev_barchart}(b) shows the variability in the number of iterations required by each technique. Qoncord achieves the highest variability in the number of iterations for all three molecules, and this variability is considerably higher than BestMap and \sol{}. Compared to \sol{}, Qoncord has 2.6$\times{}$ the variability for the \heh{} molecule, 1.6$\times{}$ the variability for the \htwo{} molecule, and 2.1$\times{}$ the variability for the \hthree{} molecule. In contrast, BestMap and \sol{} achieve similar variability for \htwo{} and \hthree{}, but BestMap has 40.9\% higher variability than \sol{} for \heh{}. \sol{} generally achieves the most stable behavior in terms of the number of iterations -- that is, it has the lowest variability in terms of convergence.

\begin{figure}[t]
    \centering
    \subfloat[]{\includegraphics[width=0.25\linewidth]{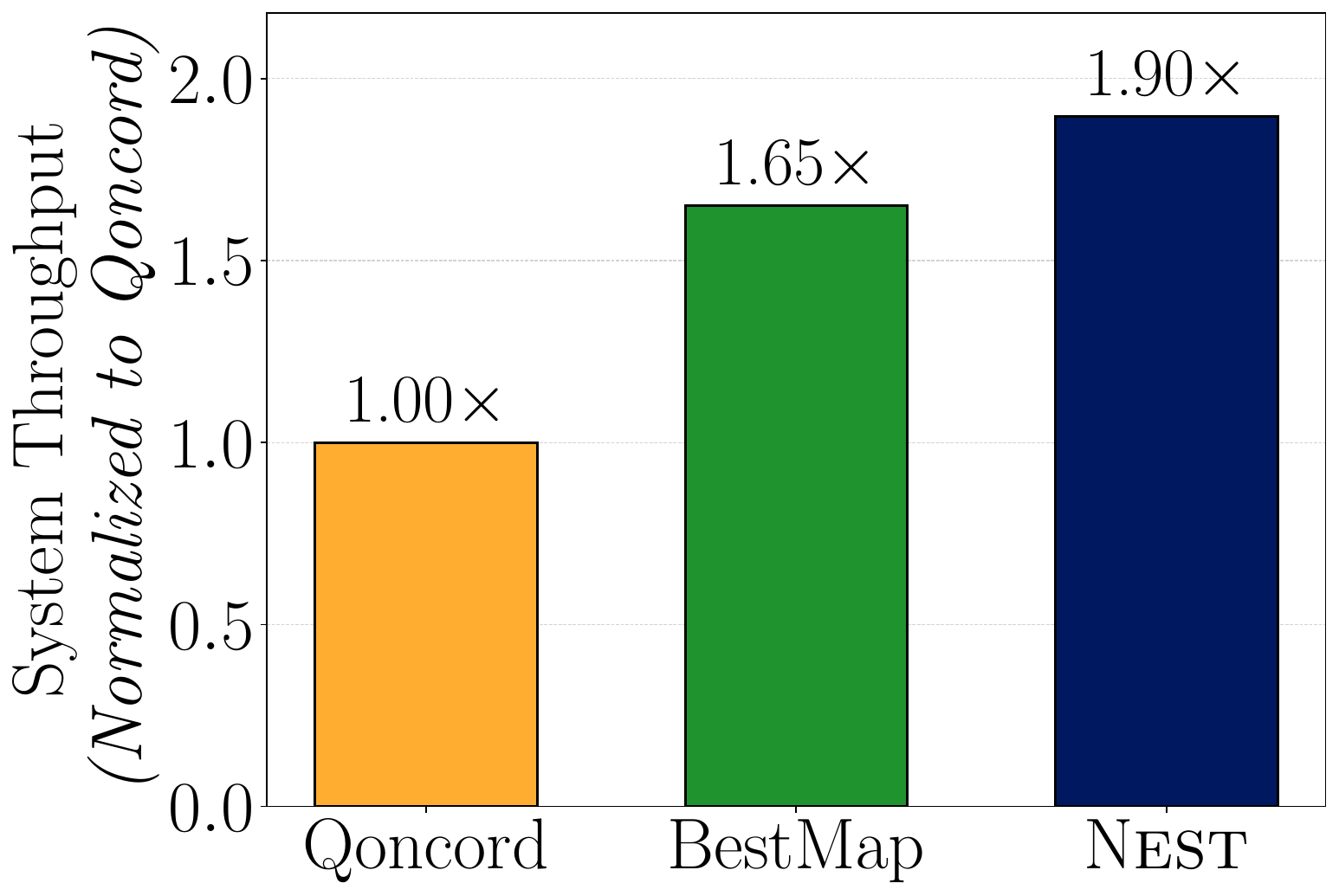}}
    \hfill
    \subfloat[]{\includegraphics[width=0.24\linewidth]{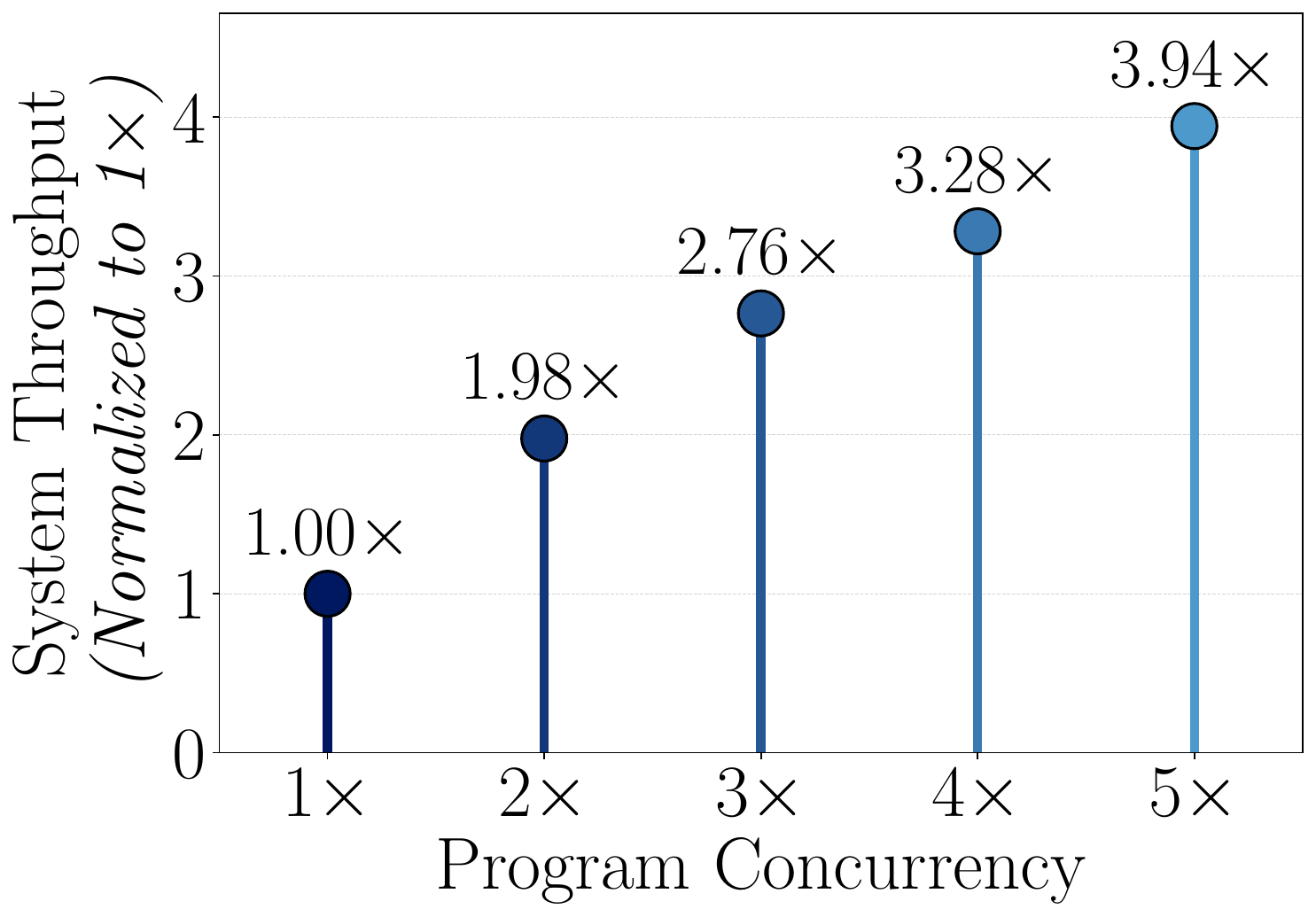}}
    \vspace{-3mm}
    \subfloat[]{\includegraphics[width=0.24\linewidth]{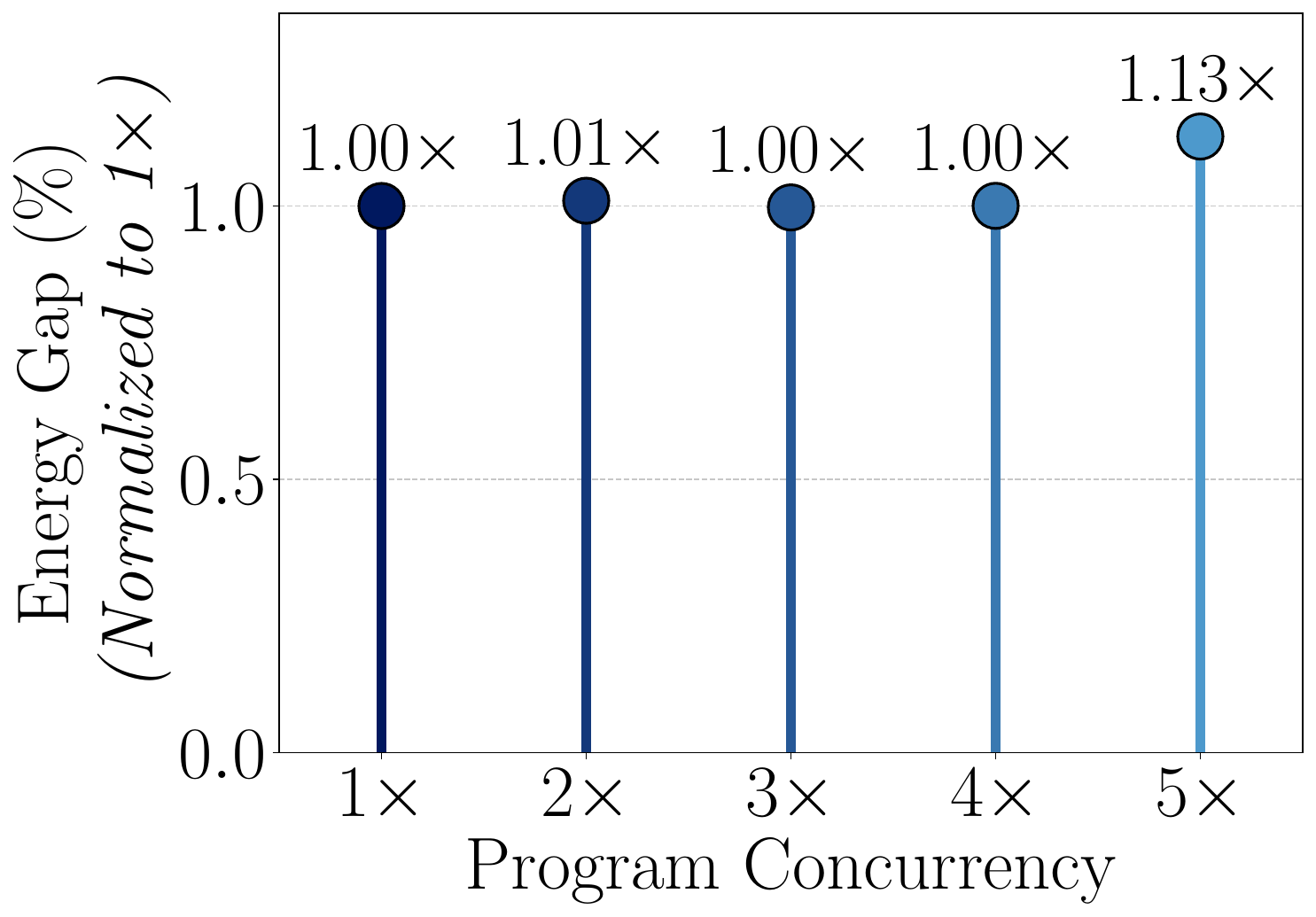}}
    \hfill
    \subfloat[]{\includegraphics[width=0.24\linewidth]{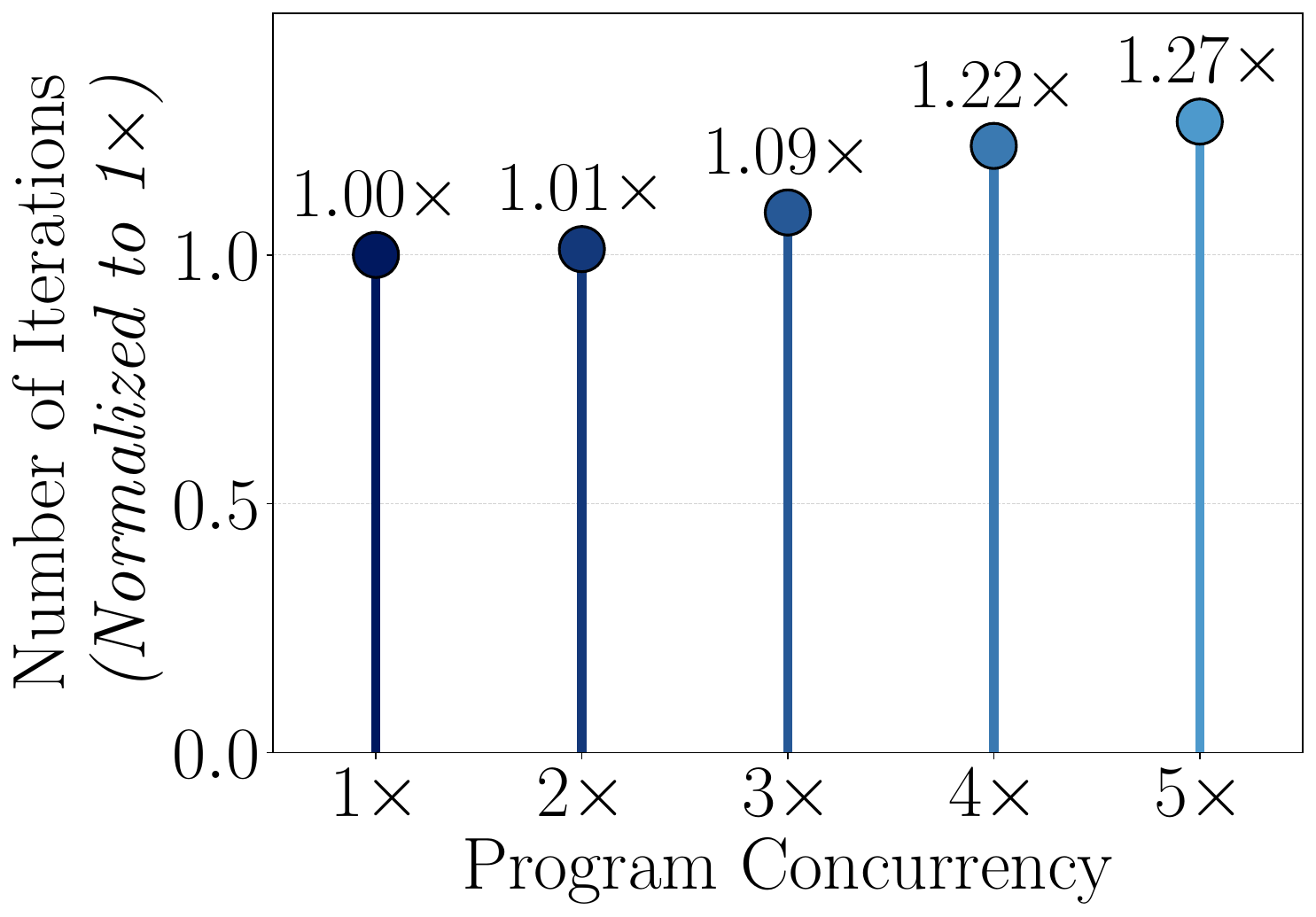}}
    \caption{(a) \sol{} achieves $1.9\times{}$ the system throughput of Qoncord and $1.2\times{}$ the system throughput of BestMap even without the deployment of \sol{}'s multi-programming. (b) Increasing the number of programs run concurrently improves the system throughput with \sol{}. (c) As \sol{} increases the program concurrency on a quantum computer, it ensures that the performance of the program is not adversely impacted. (d) The number of iterations does increase with program concurrency, but is comparable to competitive techniques.}
    \label{fig:Eval_3_inverse_iterations}
    \vspace{-4mm}
\end{figure}

\subsection{\sol{} Increases System Throughput}

We first analyze how \sol{} improves the system throughput even when it only runs one job per computer like other techniques: BestMap and Qoncord. Fig.~\ref{fig:Eval_3_inverse_iterations}(a) shows the improvement in system throughput of \sol{} compared to other competitive techniques -- normalized to the throughput of Qoncord (which achieves the lowest throughput). \sol{} achieves $1.9\times{}$ the system throughput with Qoncord and $1.15\times{}$ the system throughput with BestMap. As there is no concurrent job execution in play here, this increase in system throughput can be entirely attributed to the reduction in the number of iterations when using \sol{}, as all policies related to queuing, dispatching, and scheduling jobs are the same for all three techniques in our evaluation. \sol{}'s reduction in the number of iterations directly reduces the job runtime, allowing more jobs to run per unit of time, thus increasing the system throughput.

Next, in Fig.~\ref{fig:Eval_3_inverse_iterations}(b), we analyze how co-locating multiple programs with multi-programming can further improve the system throughput of \sol{}. As \sol{} runs more programs concurrently, it increases the system throughput approximately proportionally. With 2 concurrent programs, the system throughput increases by 2.0$\times{}$; with 3 programs, the throughput increases by 2.8$\times{}$; 4 programs lead to a 3.3$\times{}$ increase, and 5 concurrent programs can help achieve a 3.9$\times{}$ increase in the system throughput. While this increase is expected, one may wonder if it comes at the cost of VQA performance. As more programs are run concurrently, the choice of circuit maps may become more suboptimal due to the competition among the programs to be mapped to the best zone on the computer. This could, in turn, widen the energy gap and slow down convergence. However, Fig.~\ref{fig:Eval_3_inverse_iterations}(c) shows that this is not the case until a high degree of concurrency. As program concurrency is increased, the average energy gap remains the same (until a program concurrency of five is reached), indicating no decline in performance up to a concurrency of four.

Fig.~\ref{fig:Eval_3_inverse_iterations}(d) shows that as co-located program concurrency is increased, the mean number of iterations increases, indicating slower convergence. This is due to the fact that as multiple programs are co-located across the chip, the availability of suitable zones that closely match the ESP schedules decreases, thus slowing down the convergence (although the energy gap does not get affected). Nonetheless, \sol{} still performs comparable to or better than comparable techniques. Compared to \sol{} with no concurrency, \sol{} with a concurrency of 3 is 1.09$\times{}$ slower, and \sol{} with a concurrency of 5 is 1.27$\times{}$ slower. Recall that compared to \sol{} with no concurrency, BestMap is 1.13$\times{}$ slower, and Qoncord is 1.76$\times{}$ slower.

Thus, it is still better to run \sol{} with a concurrency of 5 than Qoncord (which has no concurrency). Also, it is better to run \sol{} with a concurrency of 3 than BestMap (also has no concurrency) in terms of convergence -- of course, \sol{} also has other benefits in terms of performance and system throughput. Thus, quantum cloud service providers can leverage the flexibility of \sol{} to select the concurrency level to balance the system throughput and convergence to meet user Quality of Service (QoS) expectations.

\begin{figure}[t]
    \centering
    \includegraphics[width=0.99\linewidth]{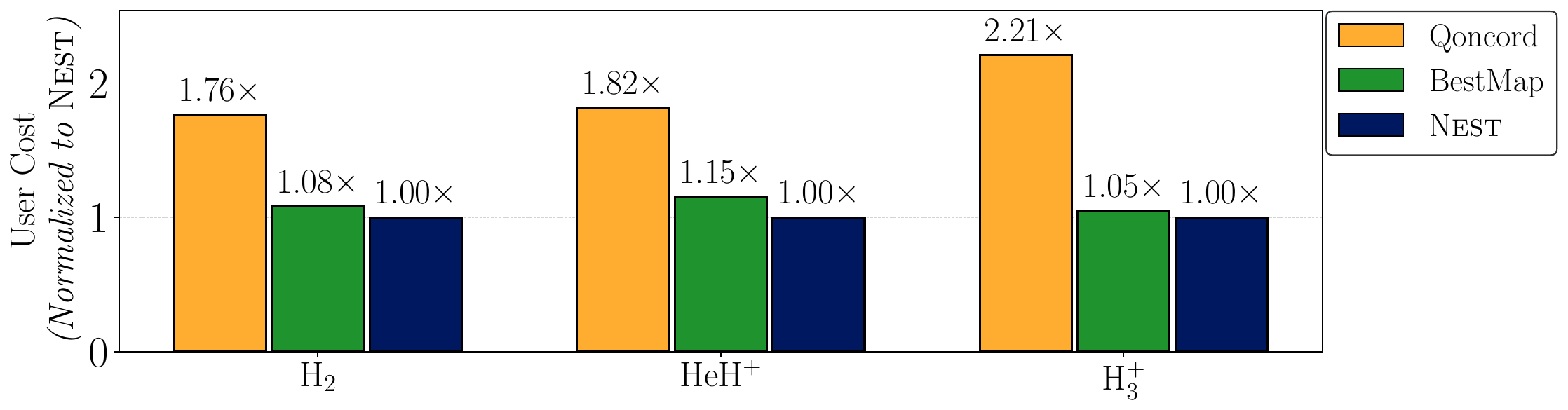}
    \vspace{-4mm}
    \caption{\sol{} helps achieve much lower user costs than competitive techniques due to its use of low-ESP maps and its need for fewer optimization iterations.}
    \label{fig:Eval_5_esp_costs}
    \vspace{-4mm}
\end{figure}

\subsection{\sol{} Decreases the Cost Incurred by Users}

We now study how \sol{} can potentially decrease the cost incurred by the users. Fig.~\ref{fig:Eval_5_esp_costs} shows the cost incurred for each molecule when users are charged based on the average ESP of the circuit maps used, the average circuit depth, and the number of algorithm iterations for different techniques. On average, users incur a 1.1$\times{}$ higher cost with BestMap and a 2.0$\times{}$ higher cost with Qoncord than with NEST. For example, users incur a 1.15$\times{}$ higher cost with BestMap and a 1.8$\times{}$ higher cost with Qoncord than with NEST for the \heh{} molecule. Our results indicate that should users be charged using our proposed metric on quantum cloud services, \sol{} can help users achieve lower cost runs, all the while achieving better performance and convergence than competitive techniques and increasing the system throughput for the cloud service providers to help them deliver as higher QoS as our earlier results demonstrate.

\subsection{\sol{}'s Compilation Times are Reasonable}

\begin{table}
    \centering
    \caption{Average compilation times of different techniques (in seconds).}
    \label{tab:compile_time}
    \vspace{-2mm}
    \renewcommand{\arraystretch}{1.0}
    \begin{tabular}{c>{\columncolor{qoncord!10}}c>{\columncolor{bestmap!10}}c>{\columncolor{nest!10}}c}
        \toprule
         & \textbf{Qoncord} & \textbf{BestMap} & \textbf{\sol{}} \\
        \midrule
           \textbf{\heh{}} & 0.12 & 12.3 & 12.4  \\
           \textbf{\htwo{}} & 0.12 & 12.3 & 12.5  \\
           \textbf{\hthree{}} & 0.14 & 13.7 & 13.8 \\
        \bottomrule
    \end{tabular}
    \vspace{-2mm}
\end{table}

All three techniques incur a one-time preexecution compilation overhead. BestMap and \sol{} first generate a specified number of mappings (described in Section~\ref{subsec: comprehensive_maps}) and compute the ESP values for each mapping. BestMap selects the mapping with the highest ESP value, while \sol{} selects the corresponding ESP according to the schedule. As shown in Table~\ref{tab:compile_time}, average compilation times for both techniques are approximately 12 seconds for \heh{} and \htwo{} molecules and 14 seconds for \hthree{} molecules. Re-mapping during qubit walk with \sol{} also requires the same low compilation times as in Table~\ref{tab:compile_time} due to the linear complexity in the number of maps with a distance of one qubit. Qoncord requires two hardware transpilations to determine the execution order of the two available computers. This process takes 0.12 seconds. 

Although Qoncord has a lower compilation time, the additional time incurred by BestMap and \sol{} enables a more comprehensive analysis of different maps. Note: the compilation times include the mapping times and are negligible compared to the execution times (order of hours), and Qoncord has especially long execution times due to more iterations.

\subsection{Scaling up \sol{} to Larger Algorithms}

\begin{figure}[t]
    \centering
    \includegraphics[width=0.99\linewidth]{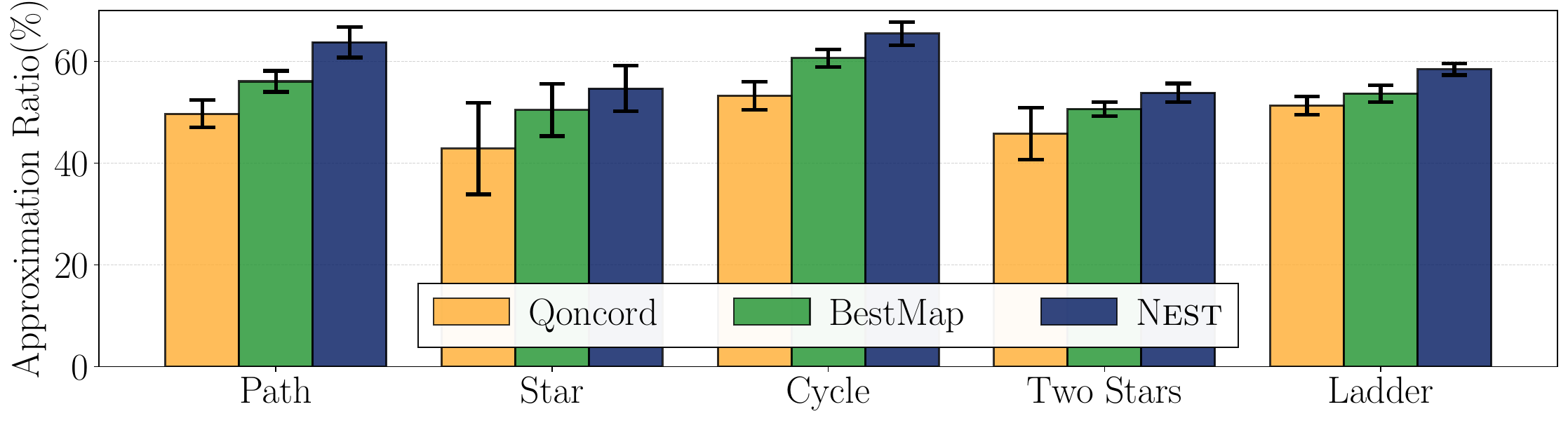}
    \vspace{-4mm}
    \caption{14-qubit MaxCut problems using QAOA with real IBM noise models. Results across five real-world instances show \sol{} consistently outperforms Qoncord and BestMap.}
    \label{fig:Eval_7_larger_circuits}
    \vspace{-4mm}
\end{figure}

We evaluated larger 14-qubit QAOA circuits with one layer (p=1) across five representative graph topologies that span diverse structural characteristics: path graphs (sequential connectivity), cycle graphs (closed loop), star graphs (centralized connectivity), two-star graphs connected by a bridge edge, and ladder graphs (parallel structure). These topologies provide varying connectivity patterns from sparse to dense configurations. All experiments include 30 runs per configuration. As shown in Fig.~\ref {fig:Eval_7_larger_circuits}, \sol{} consistently achieves higher mean approximation ratios compared to Qoncord and BestMap across all five test cases, while showing low standard deviation (indicated by the error bars). These results confirm that our approach scales effectively to larger circuits with increased qubit counts. Note that the compilation times across all techniques were also comparable and in the order of seconds.

\subsection{\sol{}'s Qubit Walk vs. Qubit Jump}

\begin{figure}[t]
    \centering
    \includegraphics[width=0.9\linewidth]{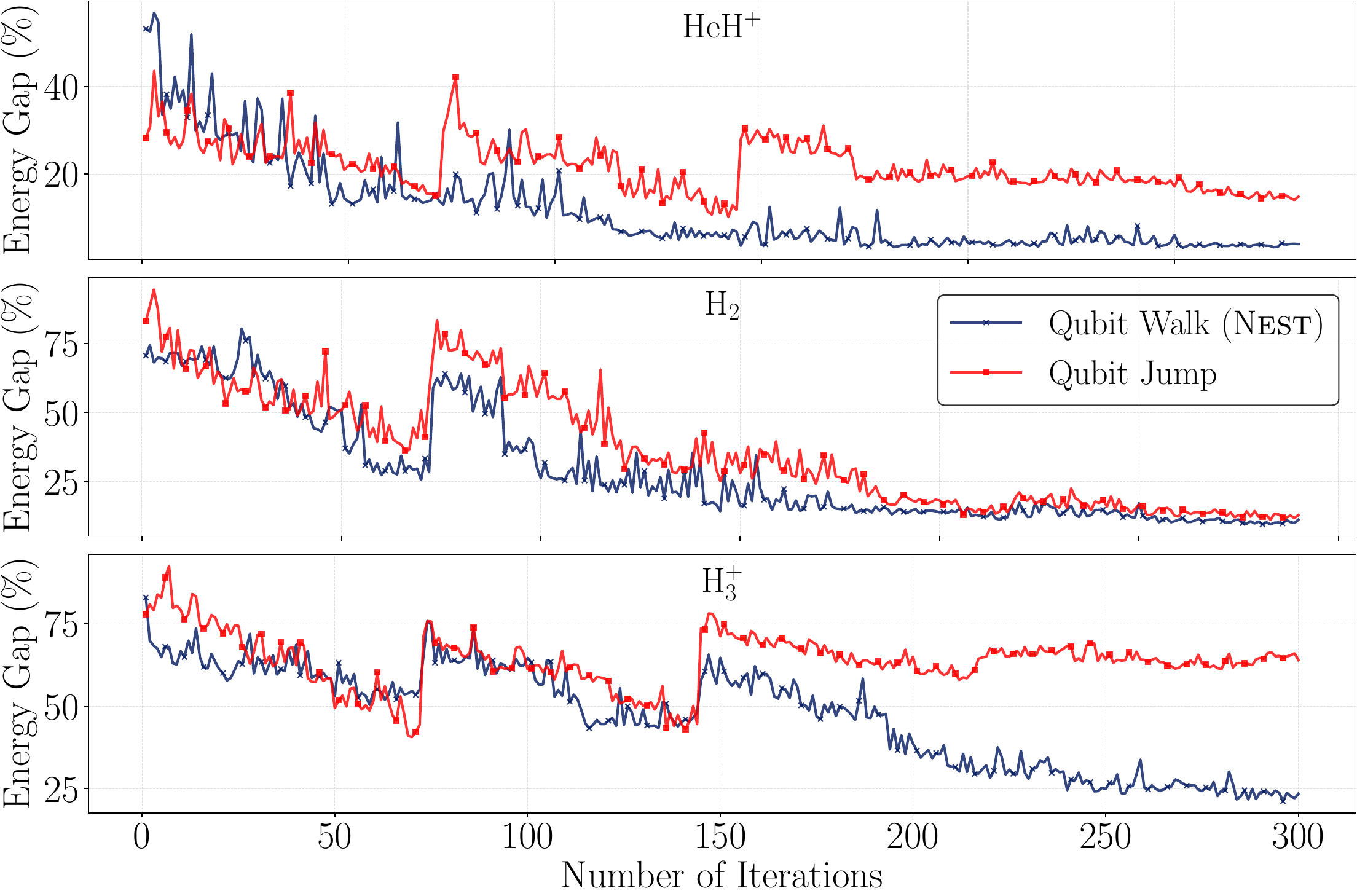}
    \vspace{-4mm}    \caption{Examples here show that Qubit jump has sharper energy surges and higher energy gaps than \sol{}.}
    \label{fig:Eval_8_quibit_jump_compare}
    \vspace{-4mm}
\end{figure}
Qubit Jump takes a direct approach by selecting any mapping closest to the target ESP. Qubit Walk (\sol{}), on the other hand, systematically explores all nearest possible mappings that remove one qubit from the current map and add another qubit from the current configuration. Energy curve optimization results from three molecular test cases (H2, H3+, HeH+) reveal that Qubit Walk significantly outperforms Qubit Jump. As shown in Fig.~\ref{fig:Eval_8_quibit_jump_compare}, Qubit Jump exhibits significant energy surges at iterations 72 and 144, where the mapping switches occur. These energy surges degrade the overall performance and hinder the circuit from finding the optimal ground state energy, resulting in larger energy gaps.

\subsection{\sol{}'s Hyperparameter Ablation}

Finally, we ablate on the hyperparameters used by \sol{}. Table~\ref{tab:ablation_study} shows how varying the number of cycles used by \sol{} and the number of iterations per cycle impact the performance and convergence of \sol{}. Recall that one circuit map is used for all iterations within one cycle, and different ones are used across different cycles.  While the energy gap is not significantly impacted by cycle count, six cycles require the fewest number of total iterations. Note that the two-cycle configuration is similar to the ESP schedule of Qoncord, as it just indicates a low-ESP circuit map initially and a high-ESP circuit map later on. We thus find that having too few cycles or too many cycles hurts the performance, as too few cycles does not provide \sol{} with the opportunity to have enough circuit map variety, while having too many cycles can cause \sol{} to hop around the optimization space too much, thus delaying convergence and requiring more iterations. We, therefore, set the default to six cycles. %Interestingly, in terms of the number of cycles used by \sol{}, the ``sweet spot'' is achieved with six cycles, where the circuit maps are selected to increase the ESP for three cycles, and then the ESP is held constant for three cycles (Inverted ReLU ESP schedule).

\begin{table}
    \centering
    \caption{Ablation analysis using the \htwo{} molecule reveals that \sol{}'s choice of 6 cycles by default and 72 iterations per cycle by default is appropriate.}
    \label{tab:ablation_study}
    \vspace{-2mm}
    \renewcommand{\arraystretch}{0.9}
    \begin{tabular}{c>{\columncolor{purple!10}}c>{\columncolor{pink!10}}c>{\columncolor{cyan!10}}c}
        \toprule
        \textbf{Configuration} & \textbf{Value} & \textbf{\makecell{Avg.\\Energy Gap}} & \textbf{\makecell{Avg. Num.\\Iterations}} \\
        \midrule
        \multirow{5}{*}[-0.1em]{\begin{tikzpicture}
            \node[draw, rounded corners, fill=yellow!20, text width=1.8cm, align=center] {Fixed\\Iterations\\Per Cycle\\(72)};
        \end{tikzpicture}} 
            & Cycles = 2 & -0.95 & 390.3\\
        %\cmidrule{2-4}
            & Cycles = 4 & -0.98 & 336.9\\
        %\cmidrule{2-4}
            & Cycles = 6 & -1.01 & 319.2 \\
        %\cmidrule{2-4}
            & Cycles = 8 & -1.03 & 389.2\\
        %\cmidrule{2-4}
            & Cycles = 10 & -1.01 & 376.8\\
        \midrule
        \multirow{5}{*}[-0.2em]{\begin{tikzpicture}
            \node[draw, rounded corners, fill=cyan!20, text width=1.8cm, align=center] {Fixed\\Cycles\\(6)};
        \end{tikzpicture}} 
            & Iters = 56 & -1.00 & 303.1\\
        %\cmidrule{2-4}
            & Iters = 64 & -1.02 & 298.5\\
        %%\cmidrule{2-4}
            & Iters = 72 & -1.01 & 319.2\\
        %\cmidrule{2-4}
            & Iters = 80 & -1.05 & 311.9\\
        %\cmidrule{2-4}
            & Iters = 88 & -1.03 & 289.2\\
        \bottomrule
    \end{tabular}
    \vspace{-2mm}
\end{table}

On the other hand, in terms of the number of iterations per cycle, we find that this hyperparameter does not significantly affect \sol{}'s performance within a wide range and, therefore, does not require considerable tuning effort. We, therefore, set the default to 72 iterations per cycle. Note that it is not necessary that \sol{} will always execute all the cycles and iterations -- in fact, as its number of iterations shows (Fig.~\ref{fig:Eval_2_energy_iterations}), it typically does not execute $6\times{}72=432$ iterations. This is due to the termination condition of \sol{}, which terminates the optimization procedure when convergence is detected. Similar to other techniques, \sol{} terminates if it detects that the energy does not decrease more than 4\% in the previous 100 iterations on a sliding window basis. This helps it terminate before the maximum number of iterations is reached.
\section{Discussion and Limitations}
\label{sec:discussion}

\noindent\textbf{Circuit Packing Density and Crosstalk:}
\sol{} does not assume smooth ESP transitions or the absence of crosstalk. All evaluations are conducted under realistic device conditions, including abrupt fidelity discontinuities, spatial contention, and non-uniform qubit coupling. Rather than enforcing exact remap alignment, \sol{} seeks maps whose fidelity profiles best match the desired ESP at each cycle. This enables the method to remain effective under non-ideal physical layouts. As a result, our experiments in Sec.~\ref{sec:evaluation} confirm that performance remains robust even when transitions are non-smooth and high-fidelity regions are contested.

\vspace{2mm}

\noindent\textbf{Size of Algorithm vs. Size of Computer:}
When a quantum algorithm fully occupies a device, intra-device mapping flexibility becomes limited. In such cases, \sol{} can be extended to operate across multiple quantum processors. Instead of classifying machines into binary categories of low- and high-fidelity (as Qoncord does), \sol{} would treat each device’s noise profile as part of a broader ESP spectrum. Devices would then be scheduled to match successive ESP targets in a way analogous to intra-device qubit walk. This extension maintains the same abstraction and allows \sol{} to remain effective for jobs constrained by device capacity.

\vspace{2mm}

\noindent\textbf{Future Fault-Tolerant Hardware:}
While \sol{} is designed for noisy machines, its core idea of fidelity-aware dynamic mapping remains relevant in fault-tolerant regimes. Even when logical qubits are stabilized via error correction, their underlying physical qubits will have different noise rates. Identifying zones with stable thermal and coherence properties will remain important. We anticipate that future QEC schedulers will need to incorporate hardware-aware considerations, and we expect ESP-like abstractions to support such decisions (e.g., mapping to ``QEC-compatible'' patches with lower syndrome extraction error rates). \sol{} thus provides a foundation for this.

\vspace{2mm}

\noindent\textbf{Uniform-Fidelity or Non-Heterogeneous Hardware:}
\sol{} is most effective on architectures with spatial fidelity heterogeneity, such as superconducting qubits, where calibration variability and manufacturing asymmetry create significant ESP variance. On emerging systems like neutral atoms or trapped ions, where qubit properties are uniform, the marginal benefit of fidelity-aware remapping may decrease. However, even in those regimes, spatial thermal effects, laser imperfections, or crosstalk during optical addressing can still induce localized heterogeneity.

\vspace{2mm}

\noindent\textbf{Cost Model and Future Pricing Strategies:}
Current quantum cloud providers (e.g., IBM, AWS) do not charge users based on fidelity. However, we posit that fidelity-aware pricing will be essential as systems scale and become shared among multiple users. In particular, high-fidelity qubits will increasingly become prime computational real estate, especially under concurrent workloads. Our model thus aligns with the natural economics of fidelity-aware execution: users who consume longer, deeper, and higher-quality resources should incur proportionally higher costs. We do not consider market factors for our cost (we model them as a constant factor for all techniques), as that is orthogonal to our work.

\section{Related Work}
\label{sec:related_work}

In addition to state-of-the-art efforts like Qoncord~\cite{10764550}, methods targeting faster convergence of VQAs have leveraged parallelism and prior knowledge to accelerate training. Resch et al.~\cite{resch2021accelerating} extended these principles specifically to VQA circuits, executing multiple runs with different parameters in parallel to overcome the sequential iteration requirements of optimization. Distributed execution frameworks like EQC~\cite{stein2022eqc} similarly use concurrent evaluations on multiple QPUs to accelerate gradient-based optimizations for VQAs while being aware of each processor’s noise profile. Multi-programming approaches for quantum computers have gained attention to address resource underutilization and throughput challenges. Early multi-programming techniques improved hardware throughput by running circuits concurrently. Das et al.~\cite{10.1145/3352460.3358287} enabled co-execution of quantum programs to improve utilization for general circuits while partitioning qubits and scheduling measurements to limit crosstalk-induced fidelity loss. QuCloud~\cite{liu2021qucloud} splits VQA workloads across multiple devices to reduce queue latency of multi-iteration executions.

At the algorithm level, techniques such as circuit cutting and parameter reuse improve the algorithmic performance under hardware constraints. CutQC~\cite{tang2021cutqc} partitions large circuits into smaller pieces executable on limited qubit devices, and transfer-learning approaches initialize VQAs with pre-trained parameters to reach near-optimal solutions faster~\cite{galda2021transferability}. Notably, CAFQA~\cite{ravi2022cafqa} provides a ``classical simulation bootstrap'' for VQAs that uses inexpensive classical approximations to find good starting parameters.

In contrast to these efforts, \sol{} is designed to simultaneously optimize algorithmic outcome, convergence speed, and system throughput. It uses a well-designed scheduler that can run multiple VQAs in parallel with resource allocation and parameter management. This approach demonstrates that high-quality VQA solutions can be obtained quickly at scale on shared quantum hardware.

\vspace{-1mm}
\section{Conclusion}
\label{sec:conclusion}

\sol{} introduces a fidelity-aware execution strategy for variational quantum algorithms that leverages intra-device heterogeneity to improve quantum program outcomes. By dynamically varying the circuit mapping using an Inverted ReLU ESP schedule, designing a structured qubit walk, and enabling multi-programming, \sol{} improves performance, accelerates convergence, and increases system throughput. Our extensive evaluation demonstrates that \sol{} converges 12.7\% faster than BestMap and 47.1\% faster than Qoncord, while reducing user cost by 1.1$\times$ and 2.0$\times$, respectively. These results highlight a simple yet powerful idea: treating fidelity as a dynamic resource unlocks new opportunities for efficient and scalable VQA execution on quantum computers.

\vspace{2mm}

\noindent\rev{\textbf{Code and Dataset Repository:} \textit{\url{https://github.com/positivetechnologylab/NEST}}.}

\section*{\rev{Acknowledgement}}

\rev{We would like to thank the anonymous reviewers and our shepherd, Professor Thirupathaiah Vasantam, for their valuable and insightful feedback that has helped improve this work. This work was supported by Rice University, the Rice University George R. Brown School of Engineering and Computing, and the Rice University Department of Computer Science. This work was supported by the DOE Quantum Testbed Finder Award DE-SC0024301, the Ken Kennedy Institute, and Rice Quantum Initiative, which is part of the Smalley-Curl Institute. This research was funded in part by: The Robert A. Welch Foundation (grant No. C-2118 A.K.); Rice University (Faculty Initiative award); NSF CAREER (award no. 2145629); an Amazon Research Award; a Microsoft Research Award. We acknowledge the use of IBM Quantum services for this work. The views expressed are those of the authors, and do not reflect the official policy or position of IBM or the IBM Quantum team.}

\balance

\bibliographystyle{ACM-Reference-Format}
\bibliography{main}

\end{document}